\documentclass[pra,twocolumn,superscriptaddress,reprint,showpacs]{revtex4-1}

\usepackage{graphicx,color}
\usepackage{amsmath}
\usepackage{amssymb}
\usepackage{epstopdf}
\usepackage[ansinew]{inputenc}
\usepackage{braket}
\usepackage{array}
\usepackage{amsfonts}
\usepackage{color}
\usepackage[colorlinks,urlcolor=blue,linkcolor=blue,citecolor=blue]{hyperref}
\usepackage{latexsym}
\usepackage{mathrsfs}
\usepackage{mathtools}
\usepackage{dsfont}
\usepackage{mathtools}
\usepackage{empheq}
\usepackage{mathbbol}
\usepackage{bbm}
\usepackage{siunitx}
\usepackage{graphicx}
\usepackage{calc}
\usepackage{comment}
\usepackage[normalem]{ulem}

\DeclarePairedDelimiter{\abs}{\lvert}{\rvert}
\newcommand{\de}{\mathrm{d}}
\newcommand{\uImm}{\text{i}}
\newcommand{\nepero}{\text{e}}

\newcommand{\Li}[1]{\mathrm{Li}_{#1}\!}
\renewcommand{\Re}{\mathop{\mathrm{Re}}}
\renewcommand{\Im}{\mathop{\text{Im}}\nolimits}
\newcommand{\uno}{\mathbb{1}}
\newcommand{\Tr}{\mathrm{Tr}\,}
\DeclareMathOperator{\arctanh}{arctanh}

\begin{document}

\title{Kitaev Chains with Long-Range Pairing} 

\author{Davide Vodola}
\affiliation{IPCMS (UMR 7504) and ISIS (UMR 7006), Universit\'{e} de Strasbourg and CNRS, 67000 Strasbourg, France}
\affiliation{Dipartimento di Fisica, Universit\`{a} di Bologna and INFN, Via Irnerio 46, 40126 Bologna, Italy}

\author{Luca Lepori}
\affiliation{IPCMS (UMR 7504) and ISIS (UMR 7006), Universit\'{e} de Strasbourg and CNRS, 67000 Strasbourg, France}

\author{Elisa Ercolessi}
\affiliation{Dipartimento di Fisica, Universit\`{a} di Bologna and INFN, Via Irnerio 46, 40126 Bologna, Italy}

\author{Alexey V. Gorshkov}
\affiliation{Joint Quantum Institute, NIST/University of Maryland, College Park,  Maryland 20742, USA}

\author{Guido Pupillo}
\affiliation{IPCMS (UMR 7504) and ISIS (UMR 7006), Universit\'{e} de Strasbourg and CNRS, 67000 Strasbourg, France}

\begin{abstract}
We propose and analyze a generalization of the Kitaev chain for fermions with long-range $p$-wave pairing, which decays with distance as a power law with exponent $\alpha$. Using the integrability of the model, we demonstrate the existence of two types of gapped regimes, where correlation functions decay exponentially at short range and algebraically at long range ($\alpha > 1$) or purely algebraically  ($\alpha < 1$). Most interestingly, along the critical lines, long-range pairing is found to break conformal symmetry for sufficiently small $\alpha$. This is accompanied by a violation of the area law for the entanglement entropy in large parts of the phase diagram in the presence of a gap, and can be detected via the dynamics of entanglement following a quench. Some of these features may be relevant for current experiments with cold atomic ions.
\end{abstract}


\maketitle

The Kitaev chain describes the dynamics of one-dimensional 
spinless fermions with superconducting $p$-wave pairing \cite{kitaev}.  Open Kitaev chains support unpaired Majorana modes exponentially localized at each end \cite{Majorana}, implying the existence of a topological superconducting phase \cite{Nayak, *Nayak2}.
Their probable recent observation in spin-orbit coupled semiconductors \cite{Franz2013, Exp,*Exp_1,*Exp_2,*Exp_3,*Exp_4} has sparked renewed interest in novel properties of topological models as well as in experimental realizations. For example, Kitaev chains with long-range hopping and pairing have been recently proposed as models for helical Shiba chains, made of magnetic impurities on an $s$-wave superconductor  \cite{Pientka2013}. 

Intimately related to the Kitaev chain, Ising-type spin chains with tunable long-range interactions can  now be realized using trapped ions coupled to motional degrees of freedom or, alternatively, using neutral atoms coupled to photonic modes~\cite{exp0,exp2,exp1,Schneider2012,Bermudez2013,*Gopalakrishnan2011, *John1990,*Shahmoon2013,*Douglas2013}. Very recently, theory and experiments have provided evidence for novel static and dynamic phenomena in these systems, such as, e.g., the non-local propagation of correlations \cite{exp2,exp1,Hauke2013,daley, vandenWorm2013} or the possible violation of the area law in 
one dimension \cite{Koffel2012}. While some of these phenomena can be explained theoretically using approximate analytical and numerical methods \cite{Koffel2012,LRB}, it remains a fundamental challenge to determine basic properties of long-range interacting systems, where methods based on short-range models may fail. 

In this work, we introduce and analyze an exactly solvable model for one-dimensional fermions with long-range pairing, decaying with distance $r$ as a power-law $\sim 1/r^\alpha$. We analyze the phase diagram as a function of the power $\alpha$ of the pairing, finding several novel features. These include: (i) gapped phases for $\alpha>1$ where the decay of correlation functions evolves from exponential to algebraic from short to long distances and (ii) a gapped phase with a purely algebraic decay of correlations for $\alpha<1$. For the open chain, we find that (iii) the localization of the edge modes, similar to the case of correlations,  
varies from hybrid (exponential followed by algebraic) for $\alpha > 1$ to purely algebraic for $\alpha < 1$, where these modes become gapped. 
Throughout the phase diagram, (iv) the entanglement entropy fails to capture some of the main features of the energy spectrum and correlation functions. However, it correctly predicts (v) an exotic transition along one of the two critical lines induced by long-range pairing  from an Ising-type theory for $\alpha >3/2$ to a Luttinger-liquid-type theory for $\alpha < 3/2$.
This corresponds to (vi) a breaking of conformal symmetry, which can be also inferred by looking at the entanglement dynamics after a quench. Finally, we discuss the relevance of these results to Ising-type chains studied in trapped-ion experiments \cite{exp2,exp1}.

We consider the following  Hamiltonian for fermionic particles on a lattice of length $L$:
\begin{eqnarray}\label{Ham}
\begin{split}
H_L &= - t \sum_{j=1}^{L} \left(a^\dagger_j a_{j+1} + \mathrm{H.c.}\right)  - \mu \sum_{j=1}^L \left(n_j - \frac{1}{2}\right) \\
&+\frac{\Delta}{2} \sum_{j=1}^L \,\sum_{\ell=1}^{L-1} d_\ell^{-\alpha} \left( a_j a_{j+\ell} + a^\dagger_{j+\ell} a^\dagger_{j}\right).
\end{split} 
\end{eqnarray}
Here, $a^{\dagger}_j$  $(a_j)$ is a fermionic creation (annihilation) operator on site $j$, $n_j=a^{\dagger}_j a_j$, and $t$ is the tunneling rate on a lattice with unit lattice constant. 
The quantities $\mu$ and $\Delta$ are the chemical potential and the strength of the fermion $p$-wave pairing, respectively. For a closed chain, we define $d_\ell = \ell$ ($d_\ell = L-\ell$) if $\ell < L/2$ ($\ell > L/2$) and choose antiperiodic boundary conditions~\footnote{Antiperiodic boundary conditions ($a_{j+L} = -a_{j}$) avoid cancellations between terms like $a_i a_{j}$ and $a_{j}a_{i+L}$ and preserve translational invariance.}. For  an open chain, we define $d_\ell = \ell$ and drop terms containing $a_{j >L}$.  Without loss of generality, we set $\Delta=2t=1$ \footnote{Different values of \(\Delta/t\) just rescale the Fermi velocity.}.

Hamiltonian \eqref{Ham} has a rich phase diagram which, when the pairing is between nearest neighbors only, coincides---via the Jordan-Wigner transformation---with that of the $XY$ model. 
The latter is a generalization of the short-range Ising model \cite{Barouch1971} and belongs to its universality class \cite{HenkelBook,*HenkelBook2}, sharing with  it gapped ferromagnetic and paramagnetic phases for $|\mu| < 1$ and $|\mu| > 1$, respectively, separated by two critical points at $\mu = \pm 1$ \cite{muss}. Furthermore, the unitary transformation \(a_i \to (-1)^i \, a^\dagger_i\) ensures that the phase diagram is identical for $\mu>0$ and $\mu<0$.

For the long-range Ising model, recent numerical results have shown algebraic decay of correlation functions in the gapped paramagnetic phase~\cite{Koffel2012}, in agreement with results from other similar spin models \cite{cirac1,Hauke2010,Peter2012,Nebendahl2013,Wall2012}. For any finite $\alpha$, however, the Hamiltonian \eqref{Ham} is no longer connected to the $XY$ model by a Jordan-Wigner transformation, implying that their respective phase diagrams can be different. In addition, for finite $\alpha$, the transformation $a_i \to (-1)^i \, a^\dagger_i $ no longer connects  $\mu > 0$ to $\mu < 0$, meaning that  the phase diagram may no longer be symmetric across the line $\mu =0$. In the following, we determine the phase diagram of  Eq.~\eqref{Ham} by first analyzing its energy spectrum, the entanglement entropy  and the decay of correlation functions for the closed chain, and then the edge modes for the open one. 

\begin{figure}[t]
\hspace{-0.5cm}
\centering
\includegraphics[scale=0.75]{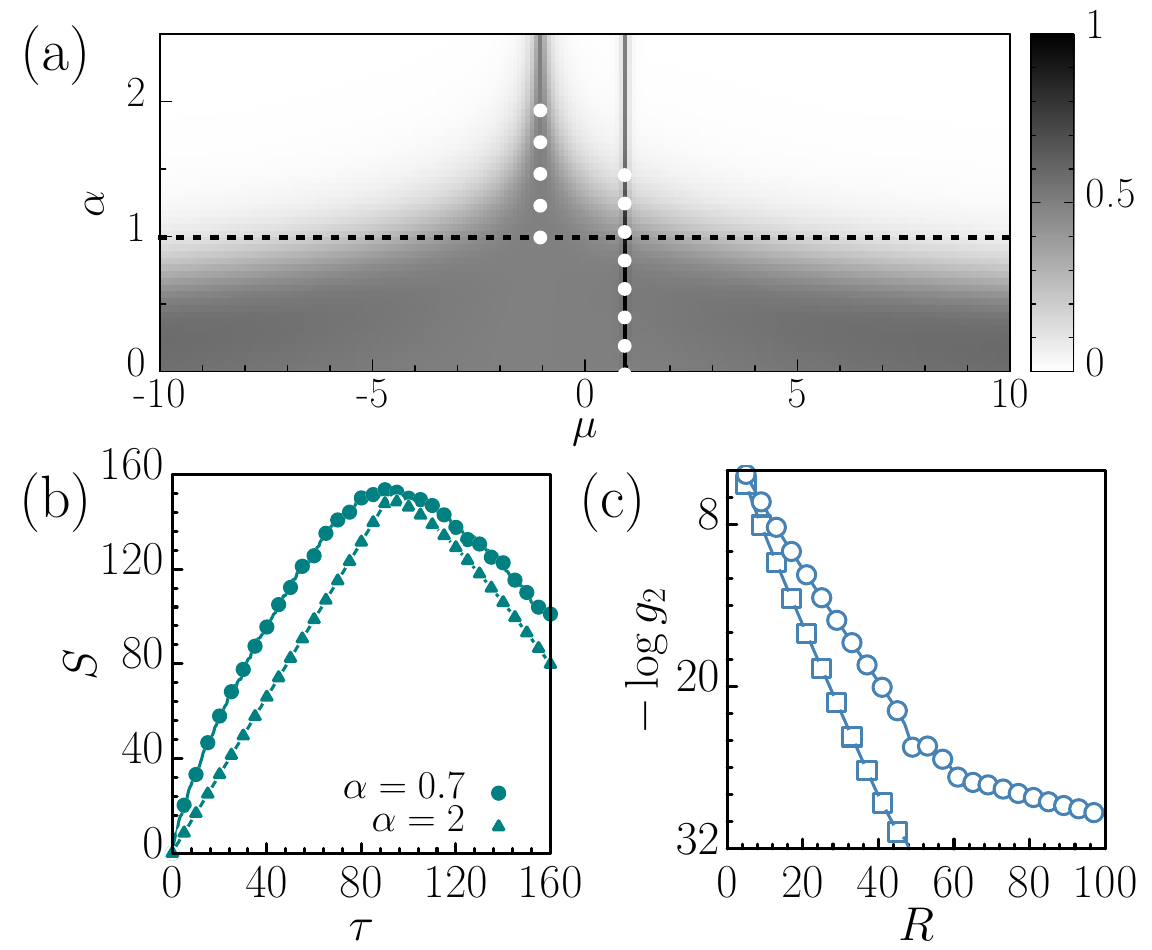}
\caption{
(a) Effective central charge $c_\mathrm{eff}$ obtained by fitting $S(L/2)$. Two gapless conformal field theories with $c=1/2$ are visible for $\mu=1$ ($\alpha>3/2$) and $\mu=-1$ ($\alpha>2$). White vertical dotted lines: gapless lines with broken conformal symmetry. 
Horizontal dashed line separates two regions: correlation functions display a hybrid exponential-algebraic ($\alpha > 1$) and purely algebraic decay ($\alpha < 1$). (b) Time evolution of $S(L/2)$ after a quench from a product state  with $\mu \gg 1$ to $\mu =1$:  $\alpha>1$, $S(L/2)$ grows  linearly, $\alpha<1$,  $S(L/2)$ grows logarithmically. (c) $g_2(R)$ correlation function  for $\mu = 2$ and $\alpha = 10$ (squares), showing exponential behavior and $\alpha=7$ (circles), showing an exponential with an algebraic tail even in the gapped region.}
\label{figure1}
\end{figure}

The {\it spectrum of excitations} is obtained via a Bogoliubov transformation as 
\begin{equation}
\lambda_{\alpha}(k_n) = \sqrt{\left(\cos k_n + \mu \right)^2 + f_{k_n , \alpha}^2} \,.
\label{eigenv}
\end{equation}
Here, $k_n\!= 2\pi (n +  1/2)/L $ are the lattice  (quasi-)momenta with $0 \leq n< L$ and the functions
$f_{k , \alpha}^{L} \equiv \sum_{l=1}^{L-1} \sin(k l)/d_\ell^\alpha$. These functions can be also evaluated in the thermodynamic limit, where they become polylogarithmic functions~\footnote{When \(L\to\infty\), $f_{k , \alpha} = \frac{1}{\uImm} \left[\Li{\alpha}(\nepero^{\uImm k})-\Li{\alpha}(\nepero^{-\uImm k})\right]$, with $\Li{\alpha}(z)$ the polylogarithmic functions \cite{ancont, ancont3, *Abramowitz1964}}. The ground state of Eq.~\eqref{Ham} is then $\Ket{\mathrm{GS}} =\prod_{n=0}^{L/2-1} \left(\cos\theta_{k_n} -\uImm \sin\theta_{k_n} a^\dagger_{k_n} a^\dagger_{-k_n} \right)\Ket{0}$, with $\tan(2\theta_{k_n}) = -f_{k_n,\alpha}/(\cos k_n + \mu)$. 

As expected from the short-range Kitaev model \cite{kitaev,muss}, Eq.~\eqref{eigenv}  is gapped for all $\alpha >1$, except for $|\mu|=1$. When $\alpha \leq 1$ the situation changes, the most evident effect being the fact that  along the 
line $\mu=-1$ the model becomes massive, as 
one can see from Eq.~\eqref{eigenv} for $k=0$. As a consequence, by tuning $\alpha$ and $\mu$, it is now possible to connect continuously the paramagnetic and ferromagnetic phases of the (short-range) Kitaev model, without closing the gap. (Without leaving the $\alpha \rightarrow \infty$ limit, such a gapped path can only be achieved with  two Kitaev wires \cite{hastings08}.) In contrast, the $\mu=1$ critical line  also survives for $\alpha \leq 1$, but the nature of the phase transition changes drastically, as we argue below. 

These features are summarized in the phase diagram of Fig.~\ref{figure1}(a). Using the method of Refs.~\cite{Peschel1987,*Peschel1989,*Peschel1999,*Peschel2012}, we compute the {\it von Neumann Entropy} 
 $S(L/2) = - \textrm{tr} \, (\rho_{L/2} \log \rho_{L/2})$, where $\rho_{L/2}$ is the reduced density matrix for half of the chain. For short-range gapped systems in one dimension,  $S(L/2)$ rapidly saturates to a constant value, a behavior known as 
 the area law~\cite{engl,fazio2008} and associated with an exponential decay of correlation functions \cite{hast}.  On the other hand, in conformally invariant models, $S(L/2)$ scales according to the formula: $S(L/2)= (c/3) \, \mathrm{log} \, L+ b$, with $b$ being a non-universal term and $c$  the central charge \cite{holz94, cal04}. In particular, $c=1/2$ for $|\mu|=1$ for the short-range Kitaev chain.

\begin{figure}[t]%
\centering\hspace{-1em}%
\includegraphics[scale=0.4]{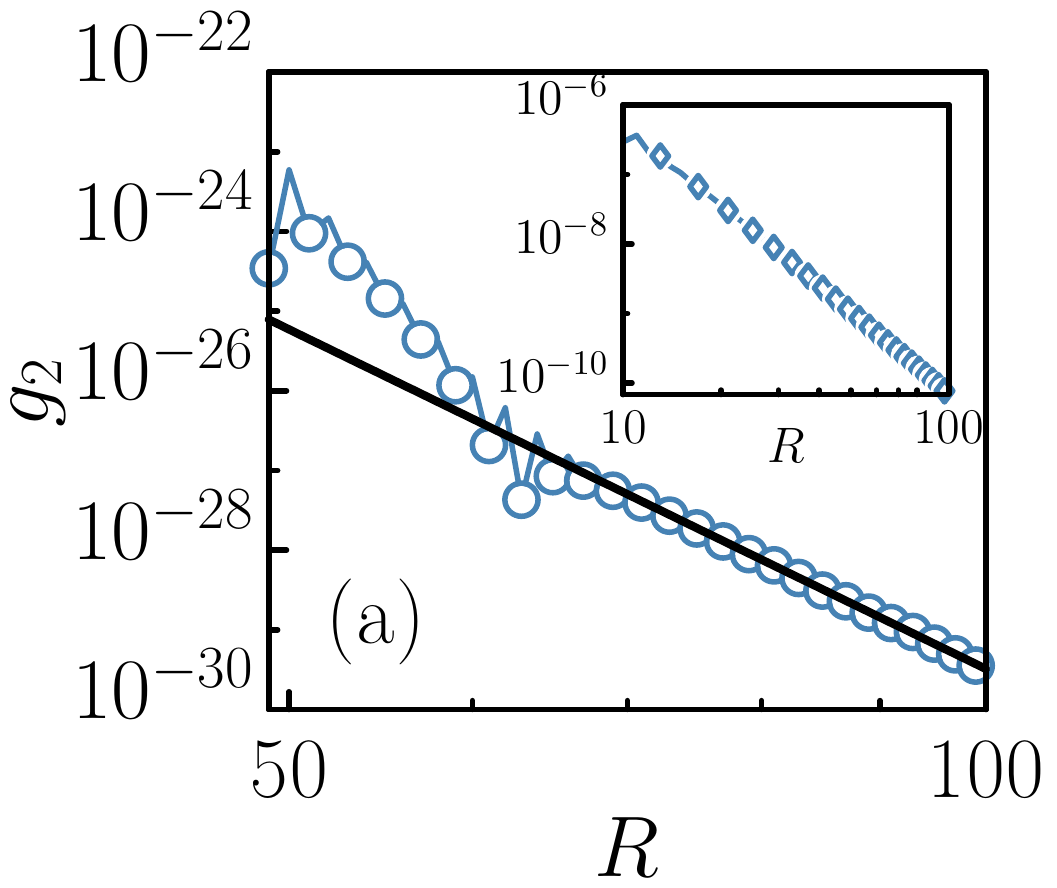}%
\includegraphics[scale=0.4]{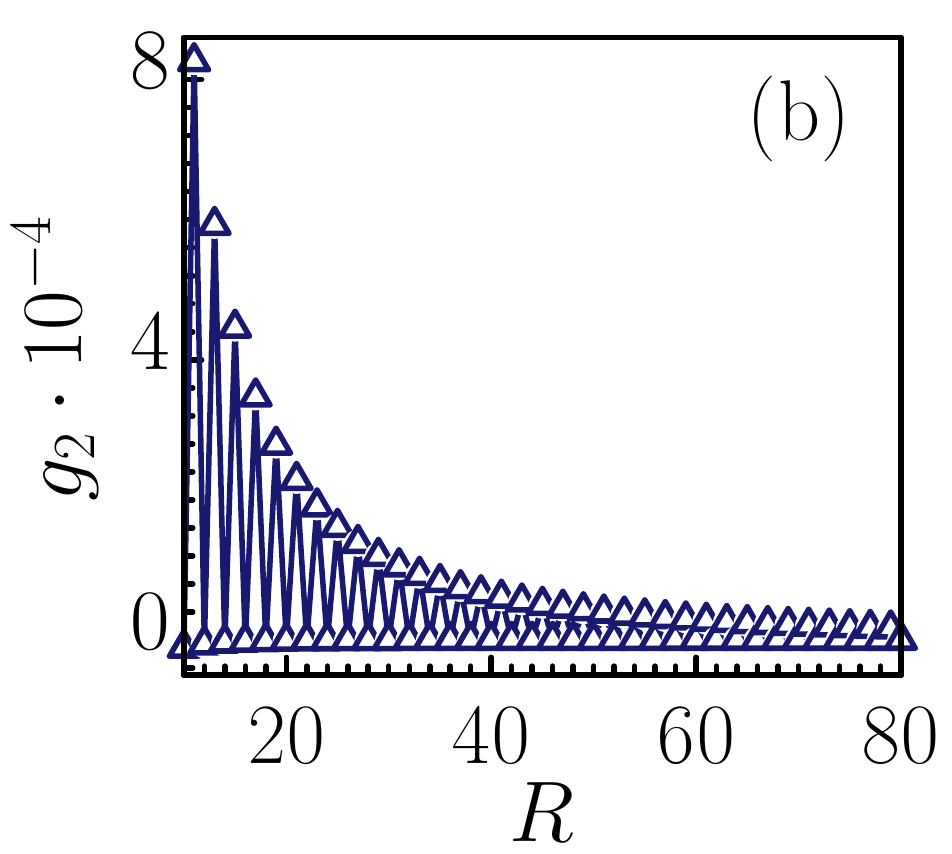}
\caption{(a) Long-distance behavior of $g_2(R)$ for $\alpha = 7$ and \(\mu=2\) in log-log scale, displaying algebraic decay.  Continuous line: analytic prediction \(g_2(R) \sim 1/R^{14}\). [Figure~\ref{figure1}(c) shows the same data set on a log plot.]
(Inset) Purely algebraic decay for $\alpha \leq 1$. Here, $\mu=2$ and $\alpha=0.5$.  (b) $g_2(R)$ for $\mu = 1$ and $\alpha = 0.5$, displaying algebraic decay with oscillating behavior.} 
\label{figure_dens}
\end{figure}

Performing finite-size scaling~\footnote{See the Supplemental Material},  we find that, surprisingly, for all $\alpha$ and $\mu$, $S(L/2)$ is well-approximated by $S(L/2) = (c_{\mathrm{\rm eff}}/3) \, \mathrm{log} \, L+ b$, where $c_{\mathrm{\rm eff}}$ is the effective central charge and is plotted in Fig.~\ref{figure1}(a).
In particular,
(i) for $\alpha > 1$, $c_{\mathrm{\rm eff}}=0$ almost everywhere in the gapped region $|\mu|\neq 1$. However, logarithmic deviations are important close to the critical line \(\mu=-1\) for $\alpha<2$ [see Fig.~\ref{figure1}(a) and below], signaling a violation of the area law.
(ii) For $\alpha < 1$, \(c_\mathrm{eff}\neq 0\) within the gapped region.
This effect is particularly evident for 
$|\mu| \lesssim 1$,  while  $c_{\mathrm{\rm eff}}=0$ for $|\mu|\to \infty $. 

In addition, most interestingly, (iii) along the critical line $\mu=1$ we observe a rapid increase of the effective central charge, obtained from the entanglement entropy formula, from $c_{\mathrm{eff}}=1/2$  {when $\alpha >3/2$} to $c_{\mathrm{\rm eff}}=1$  {when $\alpha = 0$}~\cite{Note4}. 
In a conformal field theory (CFT), the latter would correspond to a Luttinger-liquid-type theory. Indeed, we have verified numerically that density-density correlation functions [see Fig.~\ref{figure_dens}(b) and discussion below] display a strongly dimerized behavior in this region, similar to that of a charge-density wave. This peculiar behavior is further corroborated by an exact analytical computation for $\alpha=0$ \cite{Note4}. We demonstrate below that this behavior is in fact linked to the breaking of conformal symmetry below $\alpha=3/2$.

The above violations of the area law despite the presence of a gap could be naively  regarded as a failure of  $S$ to capture the physics of the model at small \(\alpha\). We recall, however, that similar behavior has been previously found for both massive quasi-free fermionic models \cite{Eisert2006} and Ising chains \cite{Koffel2012}. Thus, from a  different perspective, we may argue that 
 $S$ (together with the correlation functions which we will discuss below) is able to capture a fundamental change in the nature of the ground state, when moving towards very long-range interactions. 

For  small $\alpha$, long-range pairing becomes dominant, and its presence shows up in the physical behavior of non-local quantities, such as  $S$ and correlation functions, but cannot be inferred simply from the structure of the spectrum~\footnote{The low-lying critical finite-size spectrum, we computed, has the same degeneracy pattern as the one of the Ising model, for all $\alpha$}. At the critical line $\mu=1$, this leads to a breakdown of conformal invariance, even if the spectrum remains linear about the Fermi momentum $k_F$, as can be seen by looking at finite-size corrections to the ground-state energy density $e(\alpha)=-\sum_{n=0}^{L/2-1} \lambda_{\alpha}(k_n)/L$. The latter can be computed with the help of the Euler-MacLaurin  formula to give \cite{ancont}
\begin{equation}
e(\alpha) = e_{\infty}(\alpha) +\pi \left[\lambda_\alpha'(\pi) - \lambda_\alpha'\left(0\right)\right]/(12 L^2),
\label{scal}
\end{equation}
where \(e_{\infty}(\alpha)= - \frac{1}{\pi}\int_0^{\pi/2} \lambda_\alpha(2 x)  \, \de x\) is the value of $e(\alpha)$ in the thermodynamic limit. Exact calculations~\cite{Note4} show that, for all $\alpha > 3/2$, $\lambda'_{\alpha}(0)=0$, and thus one recovers the standard CFT result  \(e(\alpha)= e_{\infty}(\alpha) - \pi v_F \, c/(6 L^2)\), where $v_F$ is the Fermi velocity and the central charge is \(c=1/2\), in agreement with the expected value of $c$ for the short-range Ising model \cite{Belavin1984,HenkelBook,*HenkelBook2}. For $\alpha = 1$, however, the term $\lambda'_{1}(0)$ does not vanish and results in a value \(c_\mathrm{eff}\) different from that computed from the scaling of $S$ according to the formula $S(L/2) \sim (c_\mathrm{eff}/3) \, \mathrm{log} \, L$. Moreover, $\lambda'_{1}(0)$ explicitly depends on the value of the pairing coupling $\Delta$. This non-universal behavior signals a breaking of CFT and is also accompanied by the violation of the area law close to the critical line found in (i) above. Breaking of CFT is most evident for $\alpha <3/2 \, (\neq 1)$, where $\lambda'_{\alpha}(0)$ is found to diverge. A similar behavior arises at \(\mu=-1\) for \(1<\alpha<2\): the scaling \eqref{scal} fails, since the contribution from \(\lambda'_\alpha(0)\) diverges (see~Ref.~\cite{Note4}).

Following the ideas of Ref.~\cite{daley}, conformal invariance along the line  $\mu=1$ for $\alpha<1$ can also be tested by looking at the time-dependence of $S$ after a quench from $\mu \gg 1$ to  $\mu=1$. This is shown in Fig.~\ref{figure1}(b), from which it is evident that $S$ grows linearly with time $\tau$ if $\alpha >1$ as predicted by CFT \cite{Calabrese2005}, whereas it grows only logarithmically with $\tau$ when $\alpha <1$. We note that a logarithmic growth of $S$ has been recently theoretically demonstrated in Ref.~\cite{daley} for the long-range Ising chain~\cite{exp1,exp2}. We come back to this point below.

\begin{figure}[t]
\begin{center}
\includegraphics[scale=0.425]{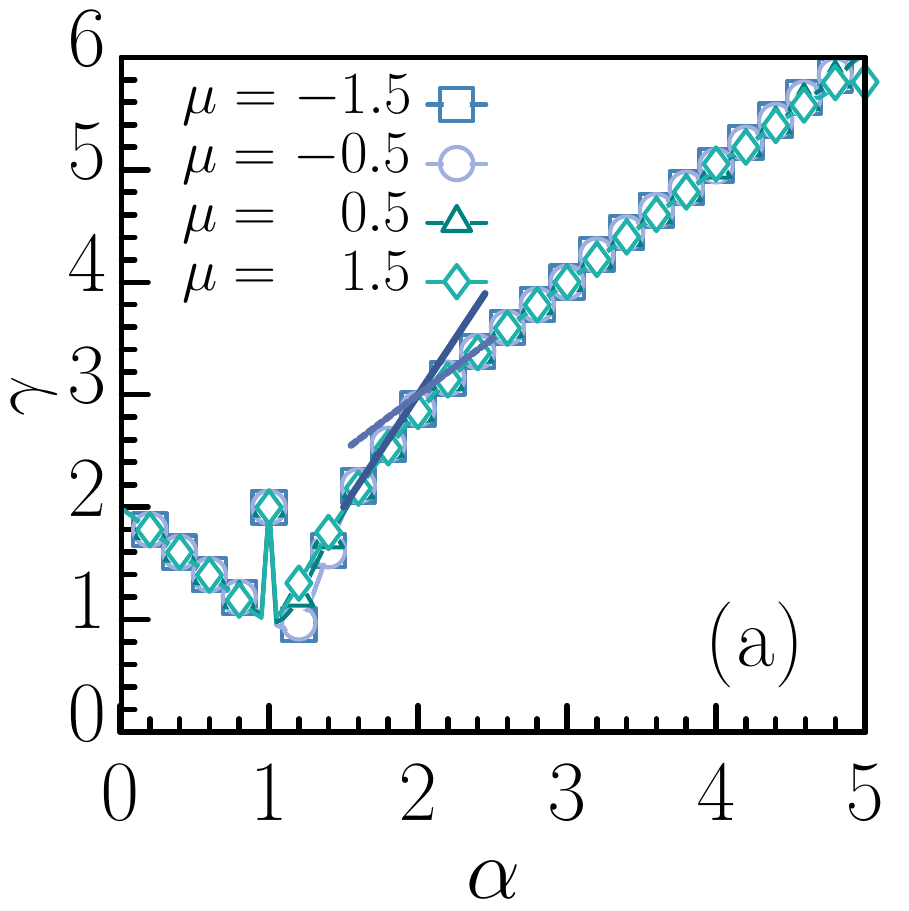}%
\includegraphics[scale=0.425]{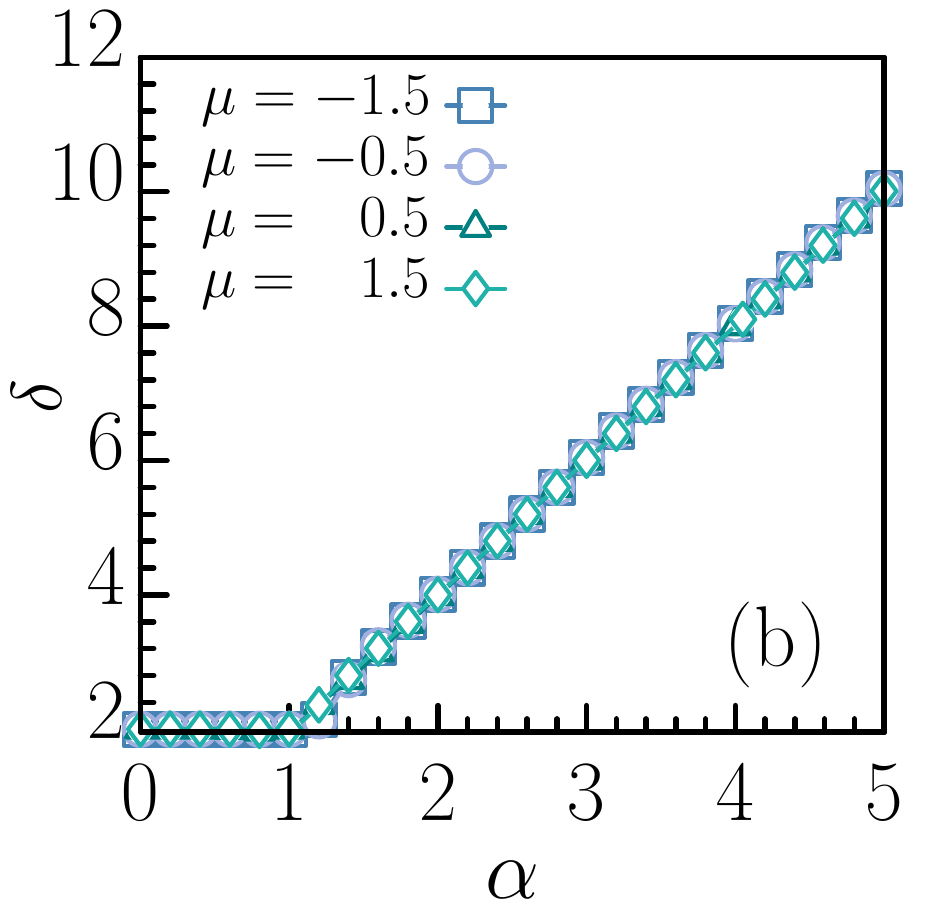}
\caption{Exponents (a) $\gamma$ and (b) $\delta$ of the algebraic decay of the one-body and two-body correlation functions vs.\ $\alpha$, obtained by fitting with power-law functions, namely, $g_1(R) \sim R^{-\gamma}$ and $g_2(R) \sim R^{-\delta}$. The equations of the two straight lines in (a) are \(2\alpha-1\) and \(\alpha+1\).}
\label{fig:Exponents}
\end{center}
\end{figure}
{\it Correlation functions} can be used to further clarify the phase diagram. The one-body correlation $g_1(|i-j|)=\langle a^\dagger_i a_j \rangle$ and the anomalous one $g_1^a(|i-j|)=\langle a^\dagger_i a^\dagger_j \rangle$ can be computed semi-analytically for finite $L$  as well as in the thermodynamic limit~\cite{Note4}. The density-density correlation  $g_2(|i-j|)=\langle n_i n_j \rangle - \langle n_i \rangle \langle n_j \rangle = {g_1^a(|i-j|)}^2-g_1(|i-j|)^2$ is then immediately obtained from Wick's theorem. Examples of \(g_2(R)\) in the regions $\alpha > 1$ and  $\alpha < 1$ are shown in Figs.\ \ref{figure1}(c) and \ref{figure_dens}(b), respectively.  In particular, Fig.\ \ref{figure1}(c) illustrates the behavior of $g_2(R)$ in the gapped phase with $\mu=2$ for $\alpha=7$ and $\alpha = 10$. While at $\alpha = 10$ the behavior seems purely exponential, similar to short-ranged gapped systems, the case $\alpha=7$ shows that the decay of $g_2(R)$ varies from an initial exponential one to an algebraic one for large $R$. This hybrid exponential-algebraic decay is consistent with the recent hybrid exponential-algebraic Lieb-Robinson bounds on the propagation of information in systems with power-law interactions \cite{LRB}. We find numerically that in our system this hybrid exponential-algebraic decay is characteristic  of \textit{all} correlation functions for finite $\alpha>1$, and obtain for the general asymptotic behavior [see Fig.~\ref{figure_dens}(a)] $g_2(R) \sim 1/R^{2\alpha}$,  $g_1(R)\sim 1/R^{\alpha+1}$ and \( g_1^a(R) \sim 1/R^\alpha\). 

These results are confirmed analytically in the thermodynamic limit, where, for example, $g_1(R)$ reads 
\begin{equation}\label{Eq:Corr}
g_1(R) = - \frac{1}{\pi}\Re \int_{0}^\pi\de k \,\nepero^{\uImm k R } \, \mathcal{C}_\alpha(k)\, ,
\end{equation}
with \(	\mathcal{C}_\alpha(k)= (\cos k + \mu)/(2\lambda_{\alpha}(k))\). Integrating by parts, one finds that the leading contribution to Eq.~\eqref{Eq:Corr}  decays as \(1/R^{n+1}\),  with \(n\) the order of the first nonvanishing odd derivative of \(\mathcal{C}_\alpha(k)\) at \(k=0\)~\footnote{For, e.g., the Ising model with  short-range interactions, contributions at $k=0$ in Eq.~\eqref{Eq:Corr} vanish}. When $\alpha >1$ is an odd integer, $n=\alpha$. A similar reasoning applies to   $g_1^a(R)$, with \(n=\alpha-1\). We finally note that the long-distance behavior of $g_2(R)$ is identical to that of the two-point correlation function of the long-range Ising chain, numerically found in \cite{Koffel2012}. Such a prediction and similar ones can be derived for this model within the spin-wave approximation~\cite{Lepori}.

The most surprising behavior occurs however for $\alpha \leq 1$, where the correlation functions  display purely algebraic decay at all length scales, as illustrated for $g_2(R)$ in Fig.\ \ref{figure_dens}(b). The fact that the behavior of the system changes when $\alpha$ falls below 1 is further illustrated in Fig.~\ref{fig:Exponents}(a). There we plot the numerically obtained exponent $\gamma$ of the algebraic decay of the single-particle correlation function $g_1(|i-j|) \sim |i-j|^{-\gamma}$ as a function of $\alpha$ at fixed $\mu$: a discontinuity occurs at $\alpha=1$ for all values of $\mu$. Similarly, Fig.~\ref{fig:Exponents}(b) shows that the scaling exponent $\delta$ of $g_2(|i-j|)\sim |i-j|^{-\delta}$  becomes $\delta=2$ for every $\alpha \leq 1$. Apart from finite-size effects that might be relevant close to $\alpha= 1$, the exponents $\gamma$ and $\delta$ are found to be independent of $\mu$. Notably the change of behavior at \(\alpha=1\) is not detected properly by \(S\). 
Finally, for the case $\mu=1$ and $\alpha \le 3/2$, integrals as in Eq.~\eqref{Eq:Corr} receive contributions from both momenta $k=0$ and $k=\pi$, resulting in the observed dimerized behavior of correlation functions [Fig.~\ref{figure_dens}(b)]~\cite{Note4}.

\begin{figure} 
\centering
\hspace{-1em}%
\includegraphics[scale=0.4]{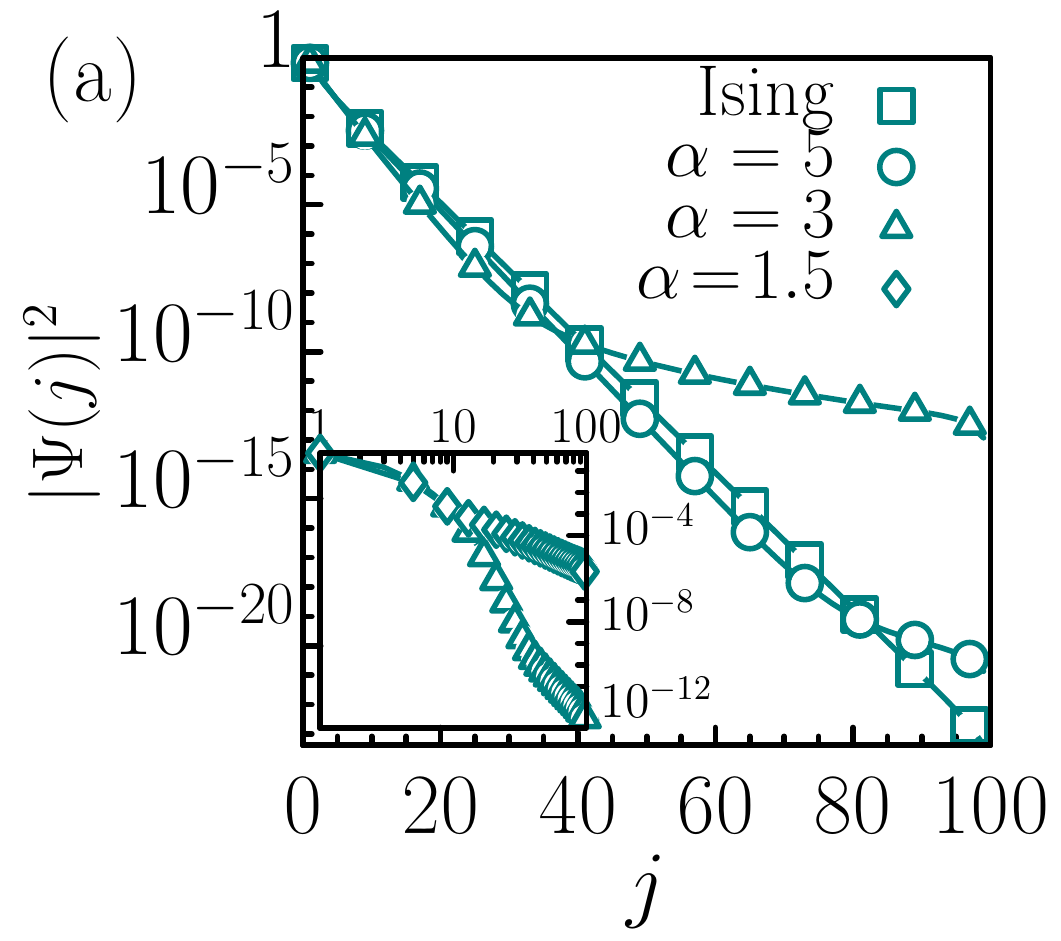}%
\includegraphics[scale=0.4]{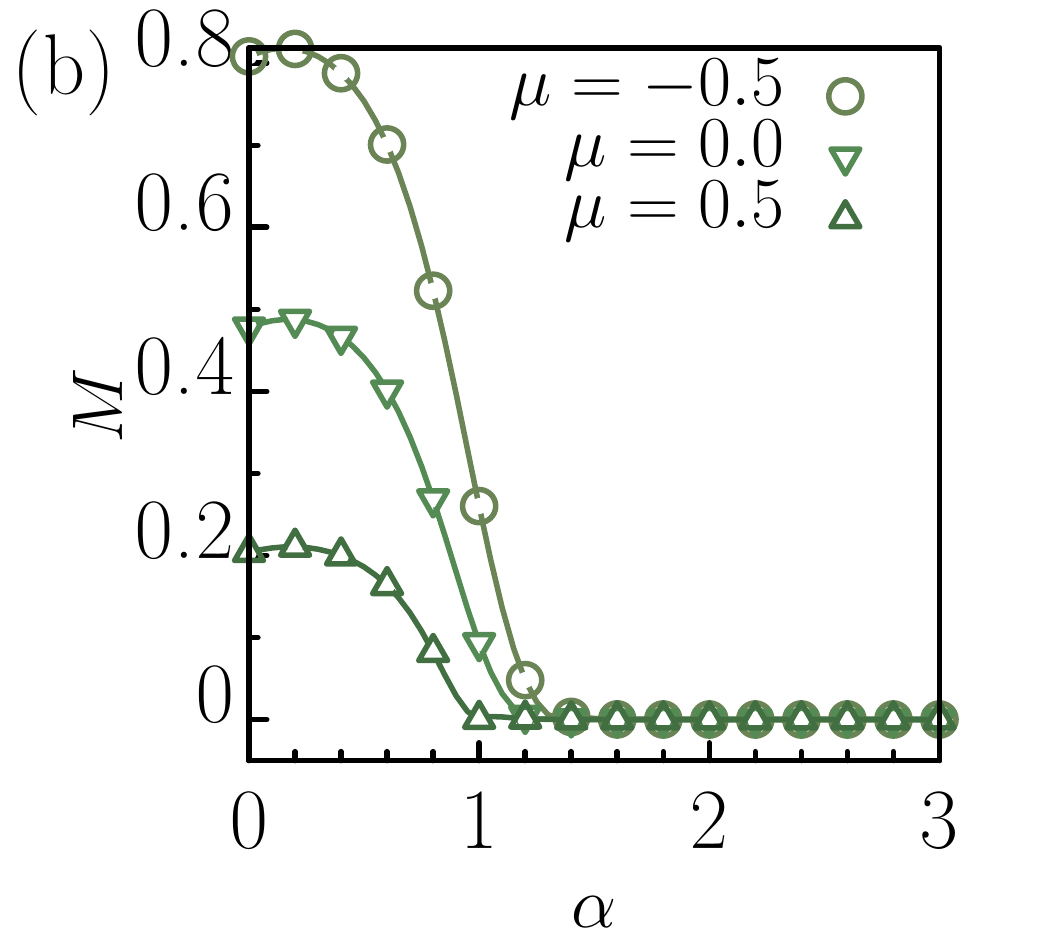}
\caption{Open chain. (a) Localization of the square of the wave function $\Psi(j)$ of the edge modes, for $\mu = 0.5$ and different $\alpha$. (b) Behavior of the mass gap $M(L\to \infty)$ for different \(\mu\) and varying \(\alpha\).}   
\label{Major}
\end{figure} 

We note that, in the thermodynamic limit of our model, we find a divergent velocity of {\it high-energy} quasiparticles for {$\alpha <3/2$ and \(\mu\neq -1\)} and for {$\alpha <2$ at \(\mu = -1\)} (more specifically at $k=0$) \cite{Note4}.  While these do not contribute to spectral properties such as the gap, they do affect the behavior of quantities such as correlation functions, entanglement entropy, and post-quench evolution~\footnote{This is, for example, at the origin of the behavior of $c_{\mathrm{eff}}$,  which is different from zero when $1<\alpha<2$ in the vicinity of the critical line $\mu=-1$}. For example, Fig.~\ref{fig:Exponents}(a) shows that the exponent $\gamma$ for the one-body correlation function changes behavior for $\alpha \lesssim 2$, with respect to the predicted value $\gamma=\alpha+1$. Related effects are also at the origin of the violation of the Lieb-Robinson bound~\cite{lieb, vandenWorm2013} recently observed in Ising-type models with long-range interactions \cite{Hauke2013, exp1, exp2}. 

{\it Open boundary conditions}.---Majorana edge modes, related to the $\mathcal{Z}_2$ symmetry of~\eqref{Ham}, arise for $\abs{\mu}<1$  if $\alpha\to \infty $ \cite{kitaev}. At finite $\alpha$, the Hamiltonian still exhibits this symmetry, and the edge modes are again expected. For $\alpha \gtrsim 1$, the decay  of the square of the edge-mode wavefunction $|\Psi(j)|^2$ ($j$ labeling the distance from an edge) mirrors  the hybrid decay of correlations discussed above [Fig.~\ref{Major}(a)]. 
A numerical fit to the algebraic tail of  $|\Psi(j)|^2$ yields $|\Psi(j)|^2 \sim j^{-2 \alpha}$ for $\alpha \gtrsim 1$, implying that $|\Psi(j)|^2$ is normalizable, as required for an edge mode \cite{Lieb1961}. 
We also note that this algebraic decay of $|\Psi(j)|^2$ is in qualitative agreement with recent calculations for helical Shiba chains~\cite{Pientka2013}. The mass $M(L)$ of the edge modes for $\alpha \gtrsim1$ exhibits similar hybrid exponential-algebraic behavior  \cite{Note4}. On the other hand, for $\alpha \lesssim 1$, $M(L)$ becomes nonzero in the limit $L \to \infty$ [Fig.~\ref{Major}(b)].

{\it Conclusions and outlook}.---In this work, we have presented and analyzed an integrable model for  
fermions with long-range pairing, finding several novel features. 
 These include gapped phases where correlation functions exhibit purely algebraic or hybrid exponential-algebraic decay. 
Moreover, for sufficiently long-range interactions, we demonstrate a breaking of the conformal symmetry along gapless lines  accompanied by a violation of  the area law in gapped phases.
It is an exciting prospect to investigate whether some of the results of the present work are in fact common to other models with long-range interactions, such as, e.g., Ising-type models with tunable interactions, as currently realized in several labs \cite{exp1,exp2}. For example, we have shown here that the breaking of conformal symmetry may be directly detected in the dynamics of the von Neumann entropy following a quench, as recently demonstrated numerically for ion chains \cite{daley}.

\medskip
We thank A.\ Turner for suggesting the idea of studying the long-range Kitaev wire. 
We thank M.\ Baranov, P.\ Calabrese, L.\ Fidkowski, M.\ Foss-Feig, Z.-X.\ Gong, F.\ Mezzacapo, S.\ Michalakis, J.\ Preskill, E.\ Rico, T.\ Roscilde, G.\ Sierra, and L.\ Taddia for useful discussions. We acknowledge support by the ERC-St Grant ColdSIM (No. 307688), EOARD, and UdS via Labex NIE and IdEX,  NSF PFC at JQI, NSF PIF, ARO, Initial Training Network COHERENCE, and computing time at the HPC-UdS.


%

\pagebreak
\widetext
\setcounter{equation}{0}
\makeatletter 
\renewcommand{\theequation}{S\@arabic\c@equation}
\makeatother

\setcounter{figure}{0}
\makeatletter 
\renewcommand{\thefigure}{S\@arabic\c@figure}
\renewcommand{\bibnumfmt}[1]{[S#1]}
\renewcommand{\citenumfont}[1]{S#1}
\makeatother

\onecolumngrid

\begin{center}
{\bf \large Supplemental Material to ``Kitaev Chains with Long-Range Pairing''}\\
\vspace{0.5cm}
{Davide Vodola,$^{1,2}$ Luca Lepori,$^{1}$ Elisa Ercolessi,$^{2}$ Alexey V. Gorshkov,$^{3}$ and Guido Pupillo$^{1}$ }\\
{$^1$}{\it IPCMS (UMR 7504) and ISIS (UMR 7006), Universit\'{e} de Strasbourg and CNRS, 67000 Strasbourg, France} \\
{$^2$}{\it Dipartimento di Fisica, Universit\`{a} di Bologna and INFN, Via Irnerio 46, 40126 Bologna, Italy} \\
{$^3$}{\it Joint Quantum Institute, NIST/University of Maryland, College Park, Maryland 20742, USA}\\
\end{center}

{\small
We present some details that were omitted in the main text. In particular, we first describe details behind analytical calculations of one-body correlation functions (Sec.~\ref{sec:corr}) and of the ground-state energy density (Secs.~\ref{sec:Velocity} and~\ref{sec:dens}). We then present details behind the scaling of the entanglement entropy (Sec.~\ref{sec:entropy})  and behind the scaling of the mass gap for the edge modes in an open chain (Sec.~\ref{sec:edge}).
}

\null\vspace{1cm}\null
\twocolumngrid

\section{Asymptotic behavior of correlation functions  \label{sec:corr}}
In this Section, we analyze the asymptotic behavior of the correlation functions of the Hamiltonian (1) of the main text.

Correlation functions take the form
\begin{gather}
\braket{a^\dag_R a_0} =- \frac{1}{\pi} \Re \int_{0}^\pi\de k \,\nepero^{\uImm k R} \, \mathcal{C}_\alpha(k),
\label{inf1} \\
\braket{a^\dag_R a^\dag_0} = \frac{1}{\pi} \Im \int_{0}^\pi\de k \,\nepero^{\uImm k R} \, \mathcal{F}_\alpha(k),
\label{inf2}
\end{gather}
with $\mathcal{C}_\alpha(k)= (\cos k + \mu)/(2\lambda_{\alpha}(k))$,  $\mathcal{F}_\alpha(k)= \Delta f_{k,\alpha}/(2\lambda_{\alpha}(k))$ and \(f_{k,\alpha}\), $\lambda_{\alpha}(k)$ as in the main text.
In order to compute the leading contribution to the integrals in Eqs.\ (\ref{inf1},\ref{inf2}) in the limit \(R\to\infty\), we will exploit the following

\emph{Theorem \cite{Sup_Fokas} -} 
Consider the integral 
\begin{equation}
I(R) = \int_{a}^{b} \de k \, f(k) \, \nepero^{\uImm k R}
\end{equation}
and assume that \(f(k)\) has \(N+1\) continuous derivatives and the $(N+2)$-th one \(f^{(N+2)}\) is piecewise continuous on \([a,b]\). Then, for \(R\rightarrow\infty\),
\begin{equation}
I(R) \simeq \sum_{n=0}^{N} \frac{(-1)^n}{(\uImm R)^{n+1}} \left[f^{(n)}(b) \, \nepero^{\uImm R b} -f^{(n)}(a) \, \nepero^{\uImm R a}  \right].
\label{Ias}
\end{equation}
For our case, $f(k)$ will be either \(\mathcal{C}_\alpha(k)\)  or \(\mathcal{F}_\alpha(k)\). In the following three subsections, we will evaluate the asymptotic behavior of $\braket{a^\dag_R a_0}$ and $\braket{a^\dag_R a^\dag_0}$ at $\alpha = 0$ and at all odd positive values of $\alpha$.

\subsection{\(\alpha=0\)}
\label{AppA}

In this case, $\lambda_{0}(k)=\sqrt{(\cos k + \mu)^2 + \Delta^2 \cot^2(k/2)}$. Then from Eqs.\ \eqref{inf1} and \eqref{Ias} the first nonvanishing contribution to \(\braket{a^\dag_R a_0}\) is given by $\mathcal{C}_{0}'(0) = (1+\mu)/(4\Delta)$, and the long range behavior of the correlator is
\begin{equation} 
\braket{a^\dag_R a_0}  = \frac{1+\mu}{4 \pi \Delta} \frac{1}{R^2} + \mathcal{O}(R^{-4}) \, .
\label{asint0}
\end{equation}

In the same way, for the anomalous correlator, Eqs.\ \eqref{inf2} and \eqref{Ias} lead to
\begin{equation}
\braket{a^\dag_R a^\dag_0} = -\frac{1}{2\pi R} + \mathcal{O}(R^{-3}).
\end{equation}

At the critical point \(\mu=1\), one has  \(\mathcal{C}'_0(0)=-\mathcal{C}'_0(\pi)=1/(2\Delta)\), so
\begin{equation}
\braket{a^\dag_R a_0} = \frac{1}{2\pi \Delta}\frac{\cos \pi R -1}{R^2}+ \mathcal{O}(R^{-4}).
\end{equation}
For the anomalous correlator, one has \(\mathcal{F}_0(0)=\mathcal{F}_0(\pi)=1/2\) and
\begin{equation}
\braket{a^\dag_R a^{\dag}_0} = \frac{1- \cos \pi R }{2 \pi R}+ \mathcal{O}(R^{-3}).
\end{equation} 
We note that the correlators here have  contributions from both points  \(k=0\) and \(k=\pi\).

Combining the previous correlators together to get the density-density correlation function \(g_2(R)\), one has
\begin{equation}
g_2(R)=\frac{1- \cos \pi R }{2 \pi^2 R^2} + \mathcal{O}(R^{-4}),
\end{equation}
which is identical to the one of a Luttinger liquid \cite{Sup_GiamarchiBook}.

\subsection{\(\alpha=1\)}

We have $\lambda_{1}(k)=\sqrt{(\cos k + \mu)^2 + \Delta^2 (\pi-k)^2}$, so from Eqs.\ \eqref{inf1} and \eqref{Ias},
\(
\mathcal{C}'_1(0) = \frac{\pi  (1+\mu )\Delta^2}{2\left(\Delta^2 \pi ^2+(1+\mu )^2\right)^{3/2}}
\)
and \(\mathcal{C}'_1(\pi) = 0, \) thus the correlation \eqref{inf1} shows a power-law decay:
\begin{equation}
\braket{a^\dag_R a_0}  =   \frac{  (1+\mu ) )\Delta^2}{2 \left(\Delta^2\pi ^2+(1+\mu )^2\right)^{3/2}} \frac{1}{R^2} + \mathcal{O}(R^{-4}).
\end{equation}

For the anomalous correlator \(\braket{a^\dag_R a^\dag_0}\) \eqref{inf2} at \(\alpha=1\), one has $\mathcal{F}_1(0) = \frac{\Delta \pi}{2\sqrt{(\mu+1)^2+\Delta^2 \pi^2}}$ and  $\mathcal{F}_1(\pi) = 0$,  
so that 
\begin{equation}
\braket{a^\dag_R a^\dag_0}  = -\frac{\Delta}{2 \sqrt{(\mu+1)^2+\Delta^2\pi^2}} \frac{1}{R}+ \mathcal{O}(R^{-3}).
\end{equation}

\subsection{Odd integer values of $\alpha>1$}
 
Using Eqs.\ \eqref{inf1} and \eqref{Ias}, one has 
\begin{equation}
\begin{split}
\braket{a^\dag_R a_0} & = -\frac{1}{\pi} \Re \int_0^\pi\de k\, \nepero^{\uImm k R} \, \mathcal{C}_\alpha(k) \\
& = \frac{1}{\pi} \sum_n \cos\left((n+1)\frac{\pi}{2}\right) \frac{\mathcal{C}_{\alpha}^{(n)}(\pi)\cos \pi R - \mathcal{C}_{\alpha}^{(n)}(0)}{R^{n+1}}.
\end{split}
\label{suma1}
\end{equation}

We need two conditions to be fulfilled to have a nonzero contribution from the sum in Eq.\ \eqref{suma1}:
\begin{itemize}
\item[(i)] \(\cos\left((n+1)\frac{\pi}{2}\right)\neq 0\), meaning that \(n\) must be odd;
\item[(ii)] either \(\mathcal{C}^{(n)}(\pi)\neq 0\) or \(\mathcal{C}^{(n)}(0)\neq 0\).
\end{itemize}

If \(\alpha\) is an odd integer \(>1\), \(\mathcal{C}^{(n)}(0)\neq 0\) if \(n \ge \alpha\), and the long-range behavior of $\braket{a^\dag_R a_0}$ is
\begin{equation}
\braket{a^\dag_R a_0}=\frac{1}{\pi}\cos\left(\frac{\pi}{2}(\alpha+1)\right) \frac{\mathcal{C}^{(\alpha)}(0)}{R^{\alpha+1}} + \mathcal{O}(R^{-(\alpha+2)}) \, .
\end{equation}

In the same way, the anomalous correlator \eqref{inf2}  is
\begin{equation}
\braket{a^\dag_R a^\dag_0}=-\frac{\Delta}{\pi}\sin\left(\frac{\pi}{2}\alpha\right) \frac{\mathcal{F}^{(\alpha)}(0)}{R^\alpha} + \mathcal{O}(R^{-(\alpha+1)}) \, .
\end{equation}

Notably we find a long-range algebraic tail for all finite $\alpha$.

\section{\label{sec:Velocity}Divergence of the quasiparticle velocity}
In this Section, we show that  \(\lambda'_{\alpha}(0)\) -- the derivative of the dispersion relation at $k=0$ -- diverges if \(\mu\neq -1\) and \(\alpha<3/2\). This can be proven by using the following expansion for the polylogarithm \cite{*[{}] [{ (eq. 25.12.12).}] Sup_ancont3}:
\begin{equation}
\Li{\alpha}(z) = \Gamma(1-\alpha)\left(\ln \frac{1}{z}\right)^{\alpha-1} + \sum_{n=0}^\infty\zeta(\alpha-n)\frac{\ln^n z}{n!}
\end{equation}
valid if \(\alpha\neq 1,2,3,\dots\) and if \(\abs{\ln z}<2\pi\). Therefore,
\begin{equation}\label{taylorPolyLog}
f_{\alpha,k}=2 \cos\left(\frac{\pi \alpha}{2}\right) \frac{\Gamma(1-\alpha)}{k^{1-\alpha}} + 2\zeta(\alpha-1) k + \mathcal{O}(k^3),
\end{equation}
and the first derivative of the dispersion relation near \(k=0\) is
\begin{equation}
\lambda'_{\alpha}(k \rightarrow 0) \sim \frac{x k+y k^{2\alpha-3}+z k^{\alpha-1}}{\sqrt{(\mu+1)^2+x' k^2 + y' k^{2\alpha-2}+z' k^\alpha}}, \label{eqn:DivergenceVelocity}
\end{equation}
with \(x,y,\dots\) coefficients that do not depend on \(k\). 

If \(\mu\neq -1\), one can see that, if \(\alpha<3/2\) and \(\alpha \neq 1\), $\lambda'_{\alpha}(k) \to \infty$ as $k \rightarrow 0$, while on the line \(\mu=-1\), \(\lambda'_{\alpha}(k)\) diverges when \(\alpha<2\) and \(\alpha\neq 1\).

\section{Ground-state energy density and central charge on the critical line $\mu = 1$ \label{sec:dens}}
In this Section, we derive the expression for the ground-state energy density given in Eq. (3) in the main text and evaluate this expression in several limits. From the Euler-MacLaurin summation formula \cite{Sup_ancont}, we find
\begin{equation}
\begin{split}
\sum_{j=0}^{n}\lambda_\alpha(a_0+j h) & = \frac{1}{2 h}\int_{a_0}^{a_0+nh}\lambda_\alpha(x) \de x \\ 
							   & + \frac{1}{2}\left(\lambda_\alpha(a_0+nh)+\lambda_\alpha(a_0)\right) \\
						       & + \frac{h}{6} \left(\lambda_\alpha'(a_0+nh)-\lambda_\alpha'(a_0) \right). 
\end{split}	
\end{equation}
In our case \(a_0=\pi/L\), \(n=L/2-1\), and \(h=2 \pi/L\), giving rise to Eq.\ (3) in the main text, which we will now use to compute the ground state energy for the cases \(\alpha>3/2\), \(\alpha=1\), and \(\alpha=0\).

If $\alpha > 3/2$, from Sec.~\ref{sec:Velocity}, one has $\lambda'_\alpha(0)=0$ and \(\lambda'_\alpha(\pi)= - v_F\), where \(v_F\) is the Fermi velocity,  
so, from Eq.\ (3) in the main text, 
\begin{equation}
e(\alpha)= e_{\infty}(\alpha) - \frac{\pi v_F c}{6 L^2}
\end{equation}
with \(c=1/2\), in agreement with the expected value of $c$ for the short-range Ising model.

If \(\alpha=1\), 
one has \(\lambda_{1}'(0)=- \frac{\pi \Delta^2}{\sqrt{4+\pi^2\Delta^2}}\),  \(\lambda_{1}'(\pi)=-\Delta\), and the Fermi velocity  \(v_F=\Delta\), so that the ground-state energy density is
\begin{equation}
e(1)= e_\infty(1) - \frac{v_F \pi}{12 L^2}\left[1-\frac{\pi\Delta}{\sqrt{4+\pi^2\Delta^2}}\right],
\end{equation}
and the effective central charge reads
\begin{equation}
c_\mathrm{eff}= \frac{1}{2}\left( 1- \frac{\pi \Delta}{\sqrt{4+\pi^2\Delta^2}}\right).
\end{equation}
The finite contribution from \(k=0\) is, in this case, not in contradiction with the results of Sec.~\ref{sec:Velocity} since the expansion \eqref{taylorPolyLog} does not hold if \(\alpha=1\).
Notably, this contribution is nonuniversal and signals a breakdown of the conformal symmetry of the model.

Finally, in the case \(\alpha=0\), from Sec.~\ref{sec:Velocity}, we have \(\lambda_{0}'(0)\to \infty\), \(\lambda_{0}'(\pi)=-\Delta/2\)  (\(v_F=\Delta/2\)) and
\begin{equation}
\begin{split}
e(0)	 &  = e_{\infty}(0) - \frac{\pi}{12 L^2} \lambda_{0}'(0) - \frac{\pi v_F}{12 L^2} \to \infty,
\end{split}
\end{equation}
differing from the $\alpha >3/2$ case because of the anomalous diverging contribution $\lambda_{0}'(0)$.

\begin{figure}[t]
\includegraphics[scale=0.4]{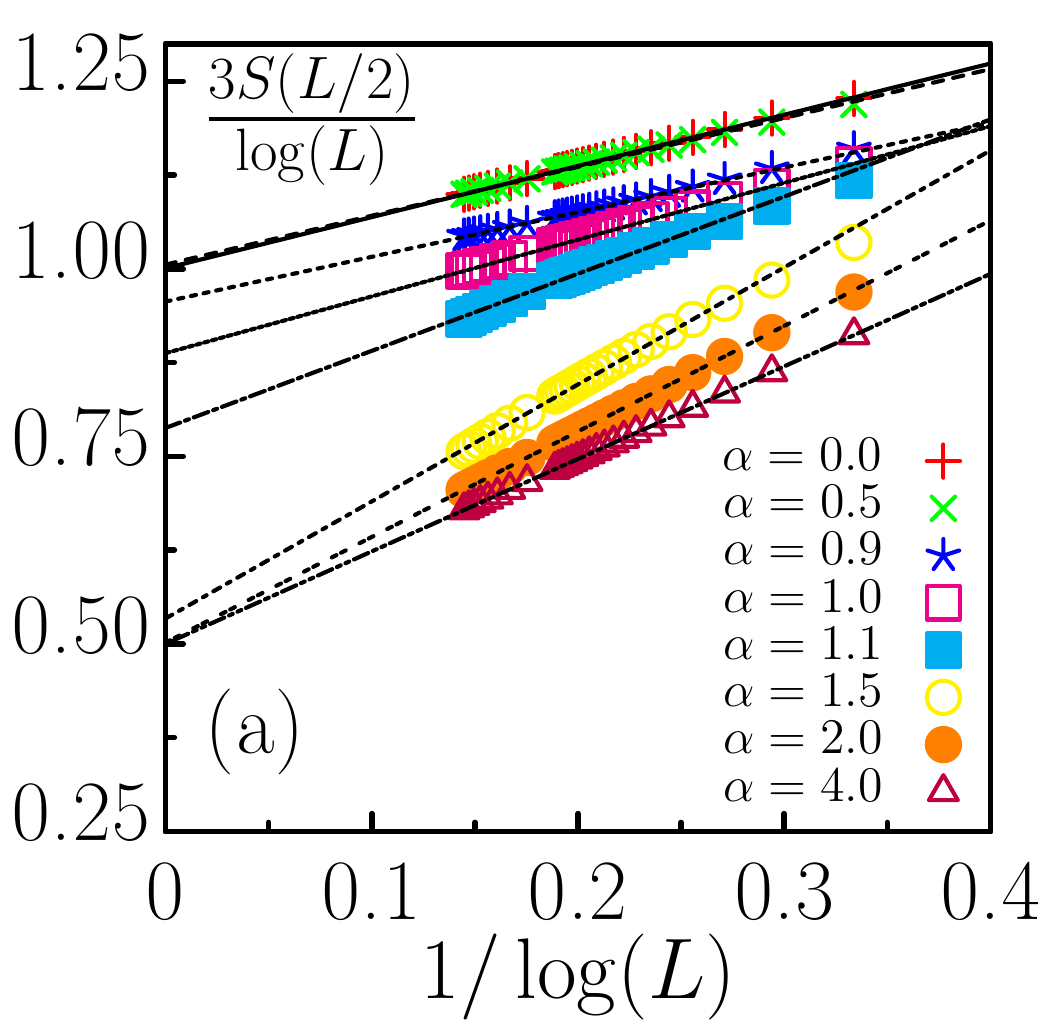}%
\includegraphics[scale=0.4]{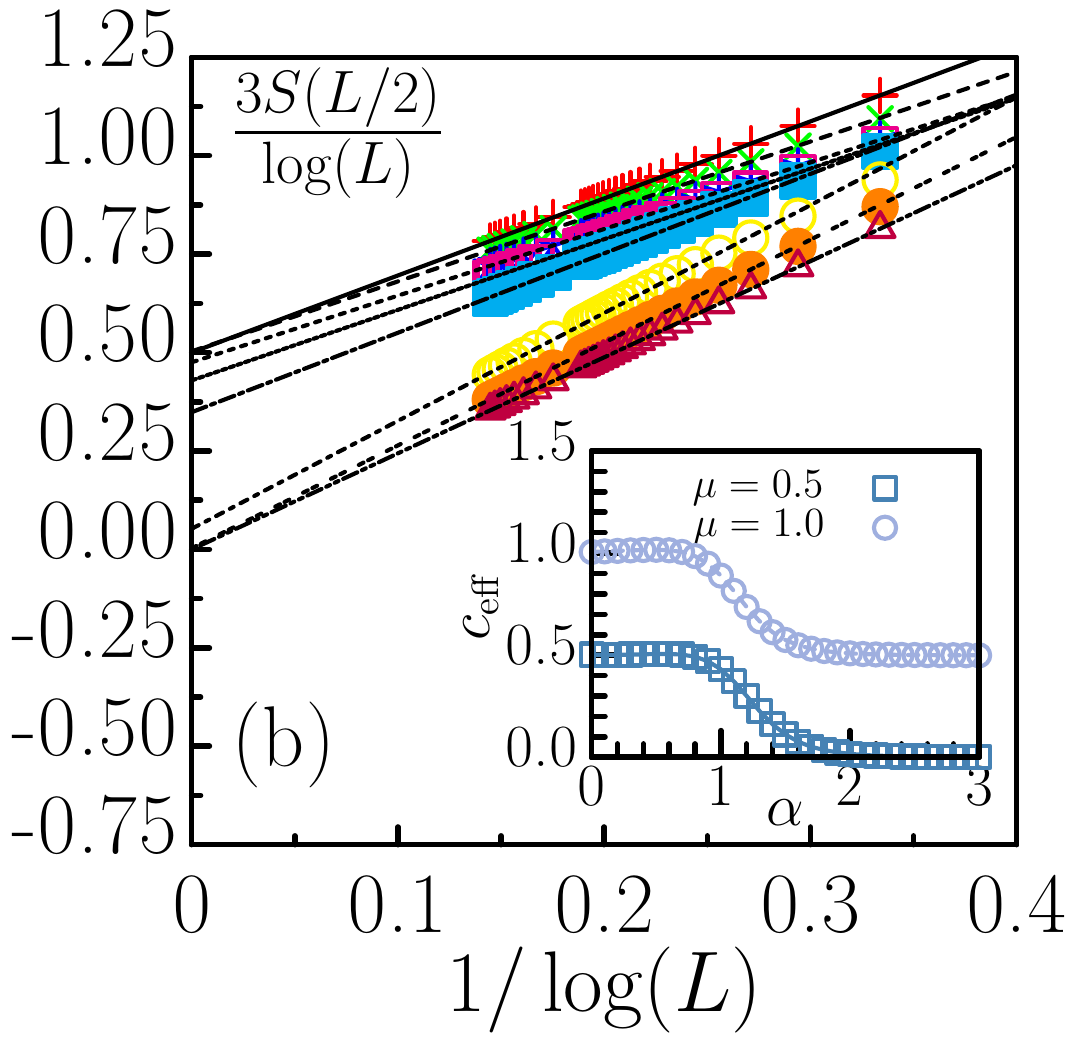}
\caption{Von Neumann entanglement entropy divided by the logarithm of the system size \(L\) vs \(1/\log(L)\) at various \(\alpha\).  The effective central charge $c_\textrm{eff}$ can be read out from the $y$-intercept. (a) \(\mu=1\) critical line. One can see that \(c_\mathrm{eff}\) tends to 1 as \(\alpha \rightarrow 0\), while $c_\mathrm{eff} = 1/2$ for \(\alpha=2\) like in the Ising model at criticality.  (b) [same symbols as in  (a)]: \(\mu=0.5\) (gapped region). \(c_\mathrm{eff}=1/2\) if \(\alpha=0\), while \(c_\mathrm{eff}=0\) if \(\alpha=2\) like in a gapped short-range system. Inset: Plot of \(c_\mathrm{eff}\) vs \(\alpha\) for \(\mu=1\) and \(\mu=0.5\).}\label{fig:EntropyScaling}
\end{figure}

\section{Entropy scaling \label{sec:entropy}}
In this Section, we present some plots illustrating the fitting procedure we used for the computation of \(c_\mathrm{eff}\). We followed the methods of Sec.~\ref{DensityMatrixFromCorrelation} to get the eigenvalues of the reduced density matrix \(\rho_{L/2}\) of half of the chain and used these eigenvalues to compute the von Neumann entropy \(S(L/2)=-\textrm{tr}( \rho_{L/2} \log\rho_{L/2})\) for different system sizes \(L\). We then fitted these values with the Cardy-Calabrese formula $S(L/2) = (c_{\mathrm{\rm eff}}/3) \, \mathrm{log} \, L+ b$ and extrapolated the thermodynamic value for \(c_{\mathrm{eff}}\). Fig.~\ref{fig:EntropyScaling} shows 
\(S(L/2)\) for different values of \(\alpha\)  for  (a)  \(\mu=1\) and (b) \(\mu=0.5\).

\begin{figure}[t!]
\centering
\includegraphics[scale=0.5]{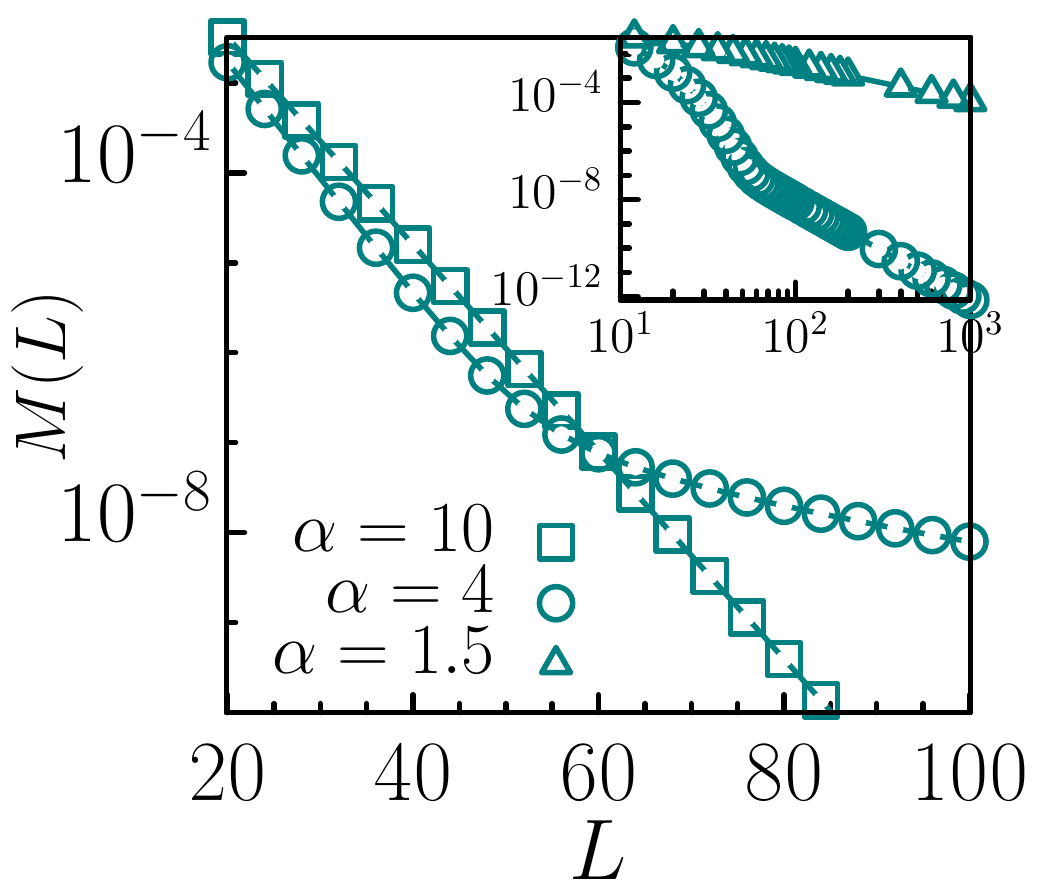}
\caption{Scaling of \(M(L)\) for  \(\mu=0.5\) and \(\alpha = 10, 4, 1.5\). The main plot shows the data in a linear-log scale, while the inset does the same in a log-log scale. }
\label{ML}
\end{figure}

\section{Scaling of the mass gap of the edge modes in an open chain \label{sec:edge}}
Hamiltonian (1) of the main text can be  written in diagonal form \(H_L=\sum_{k=1}^L \, \lambda(k) \, \eta^\dag_k \eta_k\) even with open boundary conditions following e.g.\ Refs.\ \cite{Sup_Lieb1961, Sup_Ripka}. Once  \(\lambda(k)\) are calculated, the mass gap at finite size $L$, \(M(L)\), can be easily computed as \(M(L)=\min_k{\lambda(k)}\). We show $M(L)$ in Fig.~\ref{ML}  for various $\alpha >1$. We see that, as $L$ increases, $M(L)$  falls exponentially at short distances and then algebraically at long distances. 

\section{Density matrix and entanglement entropy from correlation functions}\label{DensityMatrixFromCorrelation}
In this Section, we give some details on the technique for computing the entanglement spectrum and the von Neumann entropy for a fermionic quadratic Hamiltonian. We follow closely Ref.~\cite{Sup_Peschel1987,*Sup_Peschel1989,*Sup_Peschel1999,*Sup_Peschel2002,*Sup_Peschel2012}.\\
Consider a fermionic quadratic Hamiltonian,
\begin{equation}\label{eqn:RealHamiltonian}
H=\sum_{i,j=1}^{N} \left[c^\dagger_i t_{ij} c_j + \left(c^\dagger_i U_{ij} c^\dagger_j + \mathrm{h.c.}\right)\right]
\end{equation}
with \(t_{ij}\) (\(U_{ij}\)) a \(N\times N\)  symmetric (antisymmetric) matrix. 
Since the Hamiltonian is quadratic, Wick's theorem holds and all the correlation functions in the ground state can be expressed in terms of the one-body correlators
\begin{equation}\label{eqn:CorrelationMatrices}
C_{ij} = \braket{c^\dagger_i c_j } \qquad \qquad F_{ij} = \braket{c^\dagger_i c^\dagger_j },
\end{equation}
e.g.,
\begin{equation}\label{eqn:factorization}
\braket{c^\dag_i c^\dag_j c_k c_l} = \braket{c^\dag_i c^\dag_j}\braket{c_k c_l} -\braket{c^\dag_i c_k}\braket{c^\dag_j c_l} + \braket{c^\dag_i c_l}\braket{c^\dag_j c_k}.
\end{equation}
At the same time, if we consider a subsystem \(A\) of the whole system, the one-body correlators can be expressed by means of the reduced density matrix  \(\rho_A\) of \(A\):
\begin{equation}\label{eqn:CorrelationMatrices}
C_{ij} = \Tr{ [\rho_A c^\dagger_i c_j ]} \qquad \qquad F_{ij} = \Tr{ [\rho_A c^\dagger_i c^\dagger_j ]},
\end{equation}
while, the higher order correlations factorize as in \eqref{eqn:factorization}.
According to Wick's theorem, this property holds if the density matrix \(\rho_A\) is the exponential of a free-fermion operator \(\mathcal{H}\) \cite{Sup_Peschel2002}
\begin{equation}
\rho_A=\frac{\nepero^{-\mathcal{H}}}{Z}
\end{equation}
with 
\begin{equation}\label{eqn:hamiltonianReducedMatrix}
\mathcal{H}=\sum c^\dagger_i A_{ij} c_j + \left(c^\dagger_i B_{ij} c^\dagger_j + \mathrm{h.c.}\right).
\end{equation}
As explained in \cite{Sup_Peschel1987,*Sup_Peschel1989,*Sup_Peschel1999,*Sup_Peschel2002,*Sup_Peschel2012},
this formula implies that all the information about the density matrix is encoded in the two correlators \(C_{ij}\) and  \(F_{ij}\), easily computable once  one finds the spectrum and the ground state of \eqref{eqn:RealHamiltonian} by a suitable Bogoliubov transformation.

Indeed, denoting by \(\zeta_i\) the eingenvalues of the following matrix
\begin{equation}
W\equiv \left(C-\frac{\uno}{2} + F\right)\left(C-\frac{\uno}{2} - F\right),
\end{equation}
one can compute the eigenvalues \(\varepsilon_i\) of \(\mathcal{H}\)  as \cite{Sup_Peschel1987,*Sup_Peschel1989,*Sup_Peschel1999,*Sup_Peschel2002,*Sup_Peschel2012}:
\begin{equation}
\varepsilon_i = 2\arctanh\left(2\sqrt{\zeta_i}\right).
\end{equation}	
In this way the density matrix takes the form \(\rho_A=\otimes \rho_k\), with
\begin{equation}\label{eqn:densityMatrix}
\rho_k  =\begin{pmatrix}
(1+\nepero^{\varepsilon_k})^{-1} & 0 \\ 
0 & (1+\nepero^{-\varepsilon_k})^{-1}
\end{pmatrix},
\end{equation}
while, the von Neumann entanglement entropy reads 
\begin{equation}
S = \sum_{m} \left[\frac{\ln({1+\nepero^{\varepsilon_m}})}{1+\nepero^{\varepsilon_m}} + \frac{\ln({1+\nepero^{-\varepsilon_m}})}{1+\nepero^{-\varepsilon_m}} \right].
\end{equation}


\begin{thebibliography}{60}%
\makeatletter
\providecommand \@ifxundefined [1]{%
 \@ifx{#1\undefined}
}%
\providecommand \@ifnum [1]{%
 \ifnum #1\expandafter \@firstoftwo
 \else \expandafter \@secondoftwo
 \fi
}%
\providecommand \@ifx [1]{%
 \ifx #1\expandafter \@firstoftwo
 \else \expandafter \@secondoftwo
 \fi
}%
\providecommand \natexlab [1]{#1}%
\providecommand \enquote  [1]{``#1''}%
\providecommand \bibnamefont  [1]{#1}%
\providecommand \bibfnamefont [1]{#1}%
\providecommand \citenamefont [1]{#1}%
\providecommand \href@noop [0]{\@secondoftwo}%
\providecommand \href [0]{\begingroup \@sanitize@url \@href}%
\providecommand \@href[1]{\@@startlink{#1}\@@href}%
\providecommand \@@href[1]{\endgroup#1\@@endlink}%
\providecommand \@sanitize@url [0]{\catcode `\\12\catcode `\$12\catcode
  `\&12\catcode `\#12\catcode `\^12\catcode `\_12\catcode `\%12\relax}%
\providecommand \@@startlink[1]{}%
\providecommand \@@endlink[0]{}%
\providecommand \url  [0]{\begingroup\@sanitize@url \@url }%
\providecommand \@url [1]{\endgroup\@href {#1}{\urlprefix }}%
\providecommand \urlprefix  [0]{URL }%
\providecommand \Eprint [0]{\href }%
\providecommand \doibase [0]{http://dx.doi.org/}%
\providecommand \selectlanguage [0]{\@gobble}%
\providecommand \bibinfo  [0]{\@secondoftwo}%
\providecommand \bibfield  [0]{\@secondoftwo}%
\providecommand \translation [1]{[#1]}%
\providecommand \BibitemOpen [0]{}%
\providecommand \bibitemStop [0]{}%
\providecommand \bibitemNoStop [0]{.\EOS\space}%
\providecommand \EOS [0]{\spacefactor3000\relax}%
\providecommand \BibitemShut  [1]{\csname bibitem#1\endcsname}%
\let\auto@bib@innerbib\@empty
\bibitem [{\citenamefont {Kitaev}(2001)}]{kitaev}%
  \BibitemOpen
  \bibfield  {author} {\bibinfo {author} {\bibfnamefont {A.~Y.}\ \bibnamefont
  {Kitaev}},\ }\href {\doibase doi:10.1070/1063-7869/44/10S/S29} {\bibfield
  {journal} {\bibinfo  {journal} {Physics-Uspekhi}\ }\textbf {\bibinfo {volume}
  {44}},\ \bibinfo {pages} {131} (\bibinfo {year} {2001})}\BibitemShut
  {NoStop}%
\bibitem [{\citenamefont {Majorana}(1937)}]{Majorana}%
  \BibitemOpen
  \bibfield  {author} {\bibinfo {author} {\bibfnamefont {E.}~\bibnamefont
  {Majorana}},\ }\href {\doibase 10.1007/BF02961314} {\bibfield  {journal}
  {\bibinfo  {journal} {Il Nuovo Cimento}\ }\textbf {\bibinfo {volume} {14}},\
  \bibinfo {pages} {171} (\bibinfo {year} {1937})}\BibitemShut {NoStop}%
\bibitem [{\citenamefont {Stern}(2010)}]{Nayak}%
  \BibitemOpen
  \bibfield  {author} {\bibinfo {author} {\bibfnamefont {A.}~\bibnamefont
  {Stern}},\ }\href {\doibase doi:10.1038/nature08915} {\bibfield  {journal}
  {\bibinfo  {journal} {Nature (London)}\ }\textbf {\bibinfo {volume} {464}},\
  \bibinfo {pages} {187} (\bibinfo {year} {2010})}\BibitemShut {NoStop}%
\bibitem [{\citenamefont {Nayak}\ \emph {et~al.}(2008)\citenamefont {Nayak},
  \citenamefont {Simon}, \citenamefont {Stern}, \citenamefont {Freedman},\ and\
  \citenamefont {Das~Sarma}}]{Nayak2}%
  \BibitemOpen
  \bibfield  {author} {\bibinfo {author} {\bibfnamefont {C.}~\bibnamefont
  {Nayak}}, \bibinfo {author} {\bibfnamefont {S.~H.}\ \bibnamefont {Simon}},
  \bibinfo {author} {\bibfnamefont {A.}~\bibnamefont {Stern}}, \bibinfo
  {author} {\bibfnamefont {M.}~\bibnamefont {Freedman}}, \ and\ \bibinfo
  {author} {\bibfnamefont {S.}~\bibnamefont {Das~Sarma}},\ }\href {\doibase
  10.1103/RevModPhys.80.1083} {\bibfield  {journal} {\bibinfo  {journal} {Rev.
  Mod. Phys.}\ }\textbf {\bibinfo {volume} {80}},\ \bibinfo {pages} {1083}
  (\bibinfo {year} {2008})}\BibitemShut {NoStop}%
\bibitem [{\citenamefont {Franz}(2013)}]{Franz2013}%
  \BibitemOpen
  \bibfield  {author} {\bibinfo {author} {\bibfnamefont {M.}~\bibnamefont
  {Franz}},\ }\href {\doibase doi:10.1038/nnano.2013.33} {\bibfield  {journal}
  {\bibinfo  {journal} {Nature Nanotechnology}\ }\textbf {\bibinfo {volume}
  {8}},\ \bibinfo {pages} {149} (\bibinfo {year} {2013})}\BibitemShut {NoStop}%
\bibitem [{\citenamefont {Mourik}\ \emph {et~al.}(2012)\citenamefont {Mourik},
  \citenamefont {Zuo}, \citenamefont {Frolov}, \citenamefont {Plissard},
  \citenamefont {Bakkers},\ and\ \citenamefont {Kouwenhoven}}]{Exp}%
  \BibitemOpen
  \bibfield  {author} {\bibinfo {author} {\bibfnamefont {V.}~\bibnamefont
  {Mourik}}, \bibinfo {author} {\bibfnamefont {K.}~\bibnamefont {Zuo}},
  \bibinfo {author} {\bibfnamefont {S.~M.}\ \bibnamefont {Frolov}}, \bibinfo
  {author} {\bibfnamefont {S.~R.}\ \bibnamefont {Plissard}}, \bibinfo {author}
  {\bibfnamefont {E.~P. A.~M.}\ \bibnamefont {Bakkers}}, \ and\ \bibinfo
  {author} {\bibfnamefont {L.~P.}\ \bibnamefont {Kouwenhoven}},\ }\href
  {\doibase 10.1126/science.1222360} {\bibfield  {journal} {\bibinfo  {journal}
  {Science}\ }\textbf {\bibinfo {volume} {336}},\ \bibinfo {pages} {1003}
  (\bibinfo {year} {2012})}\BibitemShut {NoStop}%
\bibitem [{\citenamefont {Deng}\ \emph {et~al.}(2012)\citenamefont {Deng},
  \citenamefont {Yu}, \citenamefont {Huang}, \citenamefont {Larsson},
  \citenamefont {Caroff},\ and\ \citenamefont {Xu}}]{Exp_1}%
  \BibitemOpen
  \bibfield  {author} {\bibinfo {author} {\bibfnamefont {M.~T.}\ \bibnamefont
  {Deng}}, \bibinfo {author} {\bibfnamefont {C.~L.}\ \bibnamefont {Yu}},
  \bibinfo {author} {\bibfnamefont {G.~Y.}\ \bibnamefont {Huang}}, \bibinfo
  {author} {\bibfnamefont {M.}~\bibnamefont {Larsson}}, \bibinfo {author}
  {\bibfnamefont {P.}~\bibnamefont {Caroff}}, \ and\ \bibinfo {author}
  {\bibfnamefont {H.~Q.}\ \bibnamefont {Xu}},\ }\href {\doibase
  dx.doi.org/10.1021/nl303758w?} {\bibfield  {journal} {\bibinfo  {journal}
  {Nano Lett.}\ }\textbf {\bibinfo {volume} {12}},\ \bibinfo {pages} {6414}
  (\bibinfo {year} {2012})}\BibitemShut {NoStop}%
\bibitem [{\citenamefont {Das}\ \emph {et~al.}(2012)\citenamefont {Das},
  \citenamefont {Ronen}, \citenamefont {Most}, \citenamefont {Oreg},
  \citenamefont {Heiblum},\ and\ \citenamefont {Shtrikman}}]{Exp_2}%
  \BibitemOpen
  \bibfield  {author} {\bibinfo {author} {\bibfnamefont {A.}~\bibnamefont
  {Das}}, \bibinfo {author} {\bibfnamefont {Y.}~\bibnamefont {Ronen}}, \bibinfo
  {author} {\bibfnamefont {Y.}~\bibnamefont {Most}}, \bibinfo {author}
  {\bibfnamefont {Y.}~\bibnamefont {Oreg}}, \bibinfo {author} {\bibfnamefont
  {M.}~\bibnamefont {Heiblum}}, \ and\ \bibinfo {author} {\bibfnamefont
  {H.}~\bibnamefont {Shtrikman}},\ }\href {\doibase 10.1038/nphys2479}
  {\bibfield  {journal} {\bibinfo  {journal} {Nature Physics}\ }\textbf
  {\bibinfo {volume} {8}},\ \bibinfo {pages} {887} (\bibinfo {year}
  {2012})}\BibitemShut {NoStop}%
\bibitem [{\citenamefont {Rokhinson}\ \emph {et~al.}(2012)\citenamefont
  {Rokhinson}, \citenamefont {Liu},\ and\ \citenamefont {Furdyna}}]{Exp_3}%
  \BibitemOpen
  \bibfield  {author} {\bibinfo {author} {\bibfnamefont {L.~P.}\ \bibnamefont
  {Rokhinson}}, \bibinfo {author} {\bibfnamefont {X.}~\bibnamefont {Liu}}, \
  and\ \bibinfo {author} {\bibfnamefont {J.}~\bibnamefont {Furdyna}},\ }\href
  {\doibase 10.1038/nphys2429} {\bibfield  {journal} {\bibinfo  {journal}
  {Nature Physics}\ }\textbf {\bibinfo {volume} {8}},\ \bibinfo {pages} {795}
  (\bibinfo {year} {2012})}\BibitemShut {NoStop}%
\bibitem [{\citenamefont {Finck}\ \emph {et~al.}(2013)\citenamefont {Finck},
  \citenamefont {Van~Harlingen}, \citenamefont {Mohseni}, \citenamefont
  {Jung},\ and\ \citenamefont {Li}}]{Exp_4}%
  \BibitemOpen
  \bibfield  {author} {\bibinfo {author} {\bibfnamefont {A.~D.~K.}\
  \bibnamefont {Finck}}, \bibinfo {author} {\bibfnamefont {D.~J.}\ \bibnamefont
  {Van~Harlingen}}, \bibinfo {author} {\bibfnamefont {P.~K.}\ \bibnamefont
  {Mohseni}}, \bibinfo {author} {\bibfnamefont {K.}~\bibnamefont {Jung}}, \
  and\ \bibinfo {author} {\bibfnamefont {X.}~\bibnamefont {Li}},\ }\href
  {\doibase 10.1103/PhysRevLett.110.126406} {\bibfield  {journal} {\bibinfo
  {journal} {Phys. Rev. Lett.}\ }\textbf {\bibinfo {volume} {110}},\ \bibinfo
  {pages} {126406} (\bibinfo {year} {2013})}\BibitemShut {NoStop}%
\bibitem [{\citenamefont {Pientka}\ \emph {et~al.}(2013)\citenamefont
  {Pientka}, \citenamefont {Glazman},\ and\ \citenamefont {von
  Oppen}}]{Pientka2013}%
  \BibitemOpen
  \bibfield  {author} {\bibinfo {author} {\bibfnamefont {F.}~\bibnamefont
  {Pientka}}, \bibinfo {author} {\bibfnamefont {L.~I.}\ \bibnamefont
  {Glazman}}, \ and\ \bibinfo {author} {\bibfnamefont {F.}~\bibnamefont {von
  Oppen}},\ }\href {\doibase 10.1103/PhysRevB.88.155420} {\bibfield  {journal}
  {\bibinfo  {journal} {Phys. Rev. B}\ }\textbf {\bibinfo {volume} {88}},\
  \bibinfo {pages} {155420} (\bibinfo {year} {2013})}\BibitemShut {NoStop}%
\bibitem [{\citenamefont {Britton}\ \emph {et~al.}(2012)\citenamefont
  {Britton}, \citenamefont {Sawyer}, \citenamefont {Keith}, \citenamefont
  {Wang}, \citenamefont {Freericks}, \citenamefont {Uys}, \citenamefont
  {Biercuk},\ and\ \citenamefont {Bollinger}}]{exp0}%
  \BibitemOpen
  \bibfield  {author} {\bibinfo {author} {\bibfnamefont {J.~W.}\ \bibnamefont
  {Britton}}, \bibinfo {author} {\bibfnamefont {B.~C.}\ \bibnamefont {Sawyer}},
  \bibinfo {author} {\bibfnamefont {A.~C.}\ \bibnamefont {Keith}}, \bibinfo
  {author} {\bibfnamefont {C.-C.~J.}\ \bibnamefont {Wang}}, \bibinfo {author}
  {\bibfnamefont {J.~K.}\ \bibnamefont {Freericks}}, \bibinfo {author}
  {\bibfnamefont {H.}~\bibnamefont {Uys}}, \bibinfo {author} {\bibfnamefont
  {M.~J.}\ \bibnamefont {Biercuk}}, \ and\ \bibinfo {author} {\bibfnamefont
  {J.~J.}\ \bibnamefont {Bollinger}},\ }\href {\doibase 10.1038/nature10981}
  {\bibfield  {journal} {\bibinfo  {journal} {Nature}\ }\textbf {\bibinfo
  {volume} {484}},\ \bibinfo {pages} {489} (\bibinfo {year}
  {2012})}\BibitemShut {NoStop}%
\bibitem [{\citenamefont {Richerme}\ \emph {et~al.}(2014)\citenamefont
  {Richerme}, \citenamefont {Gong}, \citenamefont {Lee}, \citenamefont {Senko},
  \citenamefont {Smith}, \citenamefont {Foss-Feig}, \citenamefont {Michalakis},
  \citenamefont {Gorshkov},\ and\ \citenamefont {Monroe}}]{exp2}%
  \BibitemOpen
  \bibfield  {author} {\bibinfo {author} {\bibfnamefont {P.}~\bibnamefont
  {Richerme}}, \bibinfo {author} {\bibfnamefont {Z.-X.}\ \bibnamefont {Gong}},
  \bibinfo {author} {\bibfnamefont {A.}~\bibnamefont {Lee}}, \bibinfo {author}
  {\bibfnamefont {C.}~\bibnamefont {Senko}}, \bibinfo {author} {\bibfnamefont
  {J.}~\bibnamefont {Smith}}, \bibinfo {author} {\bibfnamefont
  {M.}~\bibnamefont {Foss-Feig}}, \bibinfo {author} {\bibfnamefont
  {S.}~\bibnamefont {Michalakis}}, \bibinfo {author} {\bibfnamefont {A.~V.}\
  \bibnamefont {Gorshkov}}, \ and\ \bibinfo {author} {\bibfnamefont
  {C.}~\bibnamefont {Monroe}},\ }\href {http://dx.doi.org/10.1038/nature13450}
  {\bibfield  {journal} {\bibinfo  {journal} {Nature}\ }\textbf {\bibinfo
  {volume} {511}},\ \bibinfo {pages} {198} (\bibinfo {year}
  {2014})}\BibitemShut {NoStop}%
\bibitem [{\citenamefont {Jurcevic}\ \emph {et~al.}(2014)\citenamefont
  {Jurcevic}, \citenamefont {Lanyon}, \citenamefont {Hauke}, \citenamefont
  {Hempel}, \citenamefont {Zoller}, \citenamefont {Blatt},\ and\ \citenamefont
  {Roos}}]{exp1}%
  \BibitemOpen
  \bibfield  {author} {\bibinfo {author} {\bibfnamefont {P.}~\bibnamefont
  {Jurcevic}}, \bibinfo {author} {\bibfnamefont {B.~P.}\ \bibnamefont
  {Lanyon}}, \bibinfo {author} {\bibfnamefont {P.}~\bibnamefont {Hauke}},
  \bibinfo {author} {\bibfnamefont {C.}~\bibnamefont {Hempel}}, \bibinfo
  {author} {\bibfnamefont {P.}~\bibnamefont {Zoller}}, \bibinfo {author}
  {\bibfnamefont {R.}~\bibnamefont {Blatt}}, \ and\ \bibinfo {author}
  {\bibfnamefont {C.~F.}\ \bibnamefont {Roos}},\ }\href
  {http://dx.doi.org/10.1038/nature13461} {\bibfield  {journal} {\bibinfo
  {journal} {Nature}\ }\textbf {\bibinfo {volume} {511}},\ \bibinfo {pages}
  {202} (\bibinfo {year} {2014})}\BibitemShut {NoStop}%
\bibitem [{\citenamefont {Schneider}\ \emph {et~al.}(2012)\citenamefont
  {Schneider}, \citenamefont {Porras},\ and\ \citenamefont
  {Schaetz}}]{Schneider2012}%
  \BibitemOpen
  \bibfield  {author} {\bibinfo {author} {\bibfnamefont {C.}~\bibnamefont
  {Schneider}}, \bibinfo {author} {\bibfnamefont {D.}~\bibnamefont {Porras}}, \
  and\ \bibinfo {author} {\bibfnamefont {T.}~\bibnamefont {Schaetz}},\ }\href
  {\doibase 10.1088/0034-4885/75/2/024401} {\bibfield  {journal} {\bibinfo
  {journal} {Rep. Prog. Phys.}\ }\textbf {\bibinfo {volume} {75}},\ \bibinfo
  {pages} {024401} (\bibinfo {year} {2012})}\BibitemShut {NoStop}%
\bibitem [{\citenamefont {Bermudez}\ \emph {et~al.}(2013)\citenamefont
  {Bermudez}, \citenamefont {Schaetz},\ and\ \citenamefont
  {Plenio}}]{Bermudez2013}%
  \BibitemOpen
  \bibfield  {author} {\bibinfo {author} {\bibfnamefont {A.}~\bibnamefont
  {Bermudez}}, \bibinfo {author} {\bibfnamefont {T.}~\bibnamefont {Schaetz}}, \
  and\ \bibinfo {author} {\bibfnamefont {M.~B.}\ \bibnamefont {Plenio}},\
  }\href {\doibase 10.1103/PhysRevLett.110.110502} {\bibfield  {journal}
  {\bibinfo  {journal} {Phys. Rev. Lett.}\ }\textbf {\bibinfo {volume} {110}},\
  \bibinfo {pages} {110502} (\bibinfo {year} {2013})}\BibitemShut {NoStop}%
\bibitem [{\citenamefont {Gopalakrishnan}\ \emph {et~al.}(2011)\citenamefont
  {Gopalakrishnan}, \citenamefont {Lev},\ and\ \citenamefont
  {Goldbart}}]{Gopalakrishnan2011}%
  \BibitemOpen
  \bibfield  {author} {\bibinfo {author} {\bibfnamefont {S.}~\bibnamefont
  {Gopalakrishnan}}, \bibinfo {author} {\bibfnamefont {B.~L.}\ \bibnamefont
  {Lev}}, \ and\ \bibinfo {author} {\bibfnamefont {P.~M.}\ \bibnamefont
  {Goldbart}},\ }\href {\doibase 10.1103/PhysRevLett.107.277201} {\bibfield
  {journal} {\bibinfo  {journal} {Phys. Rev. Lett.}\ }\textbf {\bibinfo
  {volume} {107}},\ \bibinfo {pages} {277201} (\bibinfo {year}
  {2011})}\BibitemShut {NoStop}%
\bibitem [{\citenamefont {John}\ and\ \citenamefont {Wang}(1990)}]{John1990}%
  \BibitemOpen
  \bibfield  {author} {\bibinfo {author} {\bibfnamefont {S.}~\bibnamefont
  {John}}\ and\ \bibinfo {author} {\bibfnamefont {J.}~\bibnamefont {Wang}},\
  }\href {\doibase 10.1103/PhysRevLett.64.2418} {\bibfield  {journal} {\bibinfo
   {journal} {Phys. Rev. Lett.}\ }\textbf {\bibinfo {volume} {64}},\ \bibinfo
  {pages} {2418} (\bibinfo {year} {1990})}\BibitemShut {NoStop}%
\bibitem [{\citenamefont {Shahmoon}\ and\ \citenamefont
  {Kurizki}(2013)}]{Shahmoon2013}%
  \BibitemOpen
  \bibfield  {author} {\bibinfo {author} {\bibfnamefont {E.}~\bibnamefont
  {Shahmoon}}\ and\ \bibinfo {author} {\bibfnamefont {G.}~\bibnamefont
  {Kurizki}},\ }\href {\doibase 10.1103/PhysRevA.87.033831} {\bibfield
  {journal} {\bibinfo  {journal} {Phys. Rev. A}\ }\textbf {\bibinfo {volume}
  {87}},\ \bibinfo {pages} {033831} (\bibinfo {year} {2013})}\BibitemShut
  {NoStop}%
\bibitem [{\citenamefont {Douglas}\ \emph {et~al.}(2013)\citenamefont
  {Douglas}, \citenamefont {Habibian}, \citenamefont {Gorshkov}, \citenamefont
  {Kimble},\ and\ \citenamefont {Chang}}]{Douglas2013}%
  \BibitemOpen
  \bibfield  {author} {\bibinfo {author} {\bibfnamefont {J.~S.}\ \bibnamefont
  {Douglas}}, \bibinfo {author} {\bibfnamefont {H.}~\bibnamefont {Habibian}},
  \bibinfo {author} {\bibfnamefont {A.~V.}\ \bibnamefont {Gorshkov}}, \bibinfo
  {author} {\bibfnamefont {H.~J.}\ \bibnamefont {Kimble}}, \ and\ \bibinfo
  {author} {\bibfnamefont {D.~E.}\ \bibnamefont {Chang}},\ }\href@noop {} {\
  (\bibinfo {year} {2013})},\ \Eprint {http://arxiv.org/abs/1312.2435}
  {arXiv:1312.2435 [quant-ph]} \BibitemShut {NoStop}%
\bibitem [{\citenamefont {Hauke}\ and\ \citenamefont
  {Tagliacozzo}(2013)}]{Hauke2013}%
  \BibitemOpen
  \bibfield  {author} {\bibinfo {author} {\bibfnamefont {P.}~\bibnamefont
  {Hauke}}\ and\ \bibinfo {author} {\bibfnamefont {L.}~\bibnamefont
  {Tagliacozzo}},\ }\href {\doibase 10.1103/PhysRevLett.111.207202?} {\bibfield
   {journal} {\bibinfo  {journal} {Phys. Rev. Lett.}\ }\textbf {\bibinfo
  {volume} {111}},\ \bibinfo {pages} {207202} (\bibinfo {year}
  {2013})}\BibitemShut {NoStop}%
\bibitem [{\citenamefont {Schachenmayer}\ \emph {et~al.}(2013)\citenamefont
  {Schachenmayer}, \citenamefont {Lanyon}, \citenamefont {Roos},\ and\
  \citenamefont {Daley}}]{daley}%
  \BibitemOpen
  \bibfield  {author} {\bibinfo {author} {\bibfnamefont {J.}~\bibnamefont
  {Schachenmayer}}, \bibinfo {author} {\bibfnamefont {B.~P.}\ \bibnamefont
  {Lanyon}}, \bibinfo {author} {\bibfnamefont {C.~F.}\ \bibnamefont {Roos}}, \
  and\ \bibinfo {author} {\bibfnamefont {A.~J.}\ \bibnamefont {Daley}},\ }\href
  {\doibase 10.1103/PhysRevX.3.031015} {\bibfield  {journal} {\bibinfo
  {journal} {Phys. Rev. X}\ }\textbf {\bibinfo {volume} {3}},\ \bibinfo {pages}
  {031015} (\bibinfo {year} {2013})}\BibitemShut {NoStop}%
\bibitem [{\citenamefont {Eisert}\ \emph {et~al.}(2013)\citenamefont {Eisert},
  \citenamefont {van~den Worm}, \citenamefont {Manmana},\ and\ \citenamefont
  {Kastner}}]{vandenWorm2013}%
  \BibitemOpen
  \bibfield  {author} {\bibinfo {author} {\bibfnamefont {J.}~\bibnamefont
  {Eisert}}, \bibinfo {author} {\bibfnamefont {M.}~\bibnamefont {van~den
  Worm}}, \bibinfo {author} {\bibfnamefont {S.~R.}\ \bibnamefont {Manmana}}, \
  and\ \bibinfo {author} {\bibfnamefont {M.}~\bibnamefont {Kastner}},\ }\href
  {\doibase 10.1103/PhysRevLett.111.260401} {\bibfield  {journal} {\bibinfo
  {journal} {Phys. Rev. Lett.}\ }\textbf {\bibinfo {volume} {111}},\ \bibinfo
  {pages} {260401} (\bibinfo {year} {2013})}\BibitemShut {NoStop}%
\bibitem [{\citenamefont {Koffel}\ \emph {et~al.}(2012)\citenamefont {Koffel},
  \citenamefont {Lewenstein},\ and\ \citenamefont {Tagliacozzo}}]{Koffel2012}%
  \BibitemOpen
  \bibfield  {author} {\bibinfo {author} {\bibfnamefont {T.}~\bibnamefont
  {Koffel}}, \bibinfo {author} {\bibfnamefont {M.}~\bibnamefont {Lewenstein}},
  \ and\ \bibinfo {author} {\bibfnamefont {L.}~\bibnamefont {Tagliacozzo}},\
  }\href {\doibase 10.1103/PhysRevLett.109.267203} {\bibfield  {journal}
  {\bibinfo  {journal} {Phys. Rev. Lett.}\ }\textbf {\bibinfo {volume} {109}},\
  \bibinfo {pages} {267203} (\bibinfo {year} {2012})}\BibitemShut {NoStop}%
\bibitem [{\citenamefont {Gong}\ \emph {et~al.}(2014)\citenamefont {Gong},
  \citenamefont {Foss-Feig}, \citenamefont {Michalakis},\ and\ \citenamefont
  {Gorshkov}}]{LRB}%
  \BibitemOpen
  \bibfield  {author} {\bibinfo {author} {\bibfnamefont {Z.-X.}\ \bibnamefont
  {Gong}}, \bibinfo {author} {\bibfnamefont {M.}~\bibnamefont {Foss-Feig}},
  \bibinfo {author} {\bibfnamefont {S.}~\bibnamefont {Michalakis}}, \ and\
  \bibinfo {author} {\bibfnamefont {A.~V.}\ \bibnamefont {Gorshkov}},\ }\href
  {\doibase 10.1103/PhysRevLett.113.030602} {\bibfield  {journal} {\bibinfo
  {journal} {Phys. Rev. Lett.}\ }\textbf {\bibinfo {volume} {113}},\ \bibinfo
  {pages} {030602} (\bibinfo {year} {2014})}\BibitemShut {NoStop}%
\bibitem [{Note1()}]{Note1}%
  \BibitemOpen
  \bibinfo {note} {Antiperiodic boundary conditions ($a_{j+L} = -a_{j}$) avoid
  cancellations between terms like $a_i a_{j}$ and $a_{j}a_{i+L}$ and preserve
  translational invariance.}\BibitemShut {Stop}%
\bibitem [{Note2()}]{Note2}%
  \BibitemOpen
  \bibinfo {note} {Different values of \(\Delta /t\) just rescale the Fermi
  velocity.}\BibitemShut {Stop}%
\bibitem [{\citenamefont {Barouch}\ and\ \citenamefont
  {McCoy}(1971)}]{Barouch1971}%
  \BibitemOpen
  \bibfield  {author} {\bibinfo {author} {\bibfnamefont {E.}~\bibnamefont
  {Barouch}}\ and\ \bibinfo {author} {\bibfnamefont {B.}~\bibnamefont
  {McCoy}},\ }\href {\doibase 10.1103/PhysRevA.3.786} {\bibfield  {journal}
  {\bibinfo  {journal} {Phys. Rev. A}\ }\textbf {\bibinfo {volume} {3}},\
  \bibinfo {pages} {786} (\bibinfo {year} {1971})}\BibitemShut {NoStop}%
\bibitem [{\citenamefont {di~Francesco}\ \emph {et~al.}(1997)\citenamefont
  {di~Francesco}, \citenamefont {Mathieu},\ and\ \citenamefont
  {Senechal}}]{HenkelBook}%
  \BibitemOpen
  \bibfield  {author} {\bibinfo {author} {\bibfnamefont {P.}~\bibnamefont
  {di~Francesco}}, \bibinfo {author} {\bibfnamefont {P.}~\bibnamefont
  {Mathieu}}, \ and\ \bibinfo {author} {\bibfnamefont {D.}~\bibnamefont
  {Senechal}},\ }\href@noop {} {\emph {\bibinfo {title} {Conformal Field
  Theory}}}\ (\bibinfo  {publisher} {Springer},\ \bibinfo {address} {New
  York},\ \bibinfo {year} {1997})\BibitemShut {NoStop}%
\bibitem [{\citenamefont {Henkel}(1999)}]{HenkelBook2}%
  \BibitemOpen
  \bibfield  {author} {\bibinfo {author} {\bibfnamefont {M.}~\bibnamefont
  {Henkel}},\ }\href@noop {} {\emph {\bibinfo {title} {Conformal Invariance and
  Critical Phenomena}}}\ (\bibinfo  {publisher} {Springer},\ \bibinfo {address}
  {New York},\ \bibinfo {year} {1999})\BibitemShut {NoStop}%
\bibitem [{\citenamefont {Mussardo}(2010)}]{muss}%
  \BibitemOpen
  \bibfield  {author} {\bibinfo {author} {\bibfnamefont {G.}~\bibnamefont
  {Mussardo}},\ }\href@noop {} {\emph {\bibinfo {title} {Statistical Field
  Theory, An Introduction to Exactly Solved Models in Statistical Physics}}}\
  (\bibinfo  {publisher} {Oxford University Press},\ \bibinfo {address} {New
  York},\ \bibinfo {year} {2010})\BibitemShut {NoStop}%
\bibitem [{\citenamefont {Deng}\ \emph {et~al.}(2005)\citenamefont {Deng},
  \citenamefont {Porras},\ and\ \citenamefont {Cirac}}]{cirac1}%
  \BibitemOpen
  \bibfield  {author} {\bibinfo {author} {\bibfnamefont {X.-L.}\ \bibnamefont
  {Deng}}, \bibinfo {author} {\bibfnamefont {D.}~\bibnamefont {Porras}}, \ and\
  \bibinfo {author} {\bibfnamefont {J.~I.}\ \bibnamefont {Cirac}},\ }\href
  {\doibase 10.1103/PhysRevA.72.063407} {\bibfield  {journal} {\bibinfo
  {journal} {Phys. Rev. A}\ }\textbf {\bibinfo {volume} {72}},\ \bibinfo
  {pages} {063407} (\bibinfo {year} {2005})}\BibitemShut {NoStop}%
\bibitem [{\citenamefont {Hauke}\ \emph {et~al.}(2010)\citenamefont {Hauke},
  \citenamefont {Cucchietti}, \citenamefont {M\"uller-Hermes}, \citenamefont
  {Cirac},\ and\ \citenamefont {Lewenstein}}]{Hauke2010}%
  \BibitemOpen
  \bibfield  {author} {\bibinfo {author} {\bibfnamefont {P.}~\bibnamefont
  {Hauke}}, \bibinfo {author} {\bibfnamefont {F.~M.}\ \bibnamefont
  {Cucchietti}}, \bibinfo {author} {\bibfnamefont {M.}~\bibnamefont
  {M\"uller-Hermes}, \bibfnamefont {A.~Ba\~nuls}}, \bibinfo {author}
  {\bibfnamefont {J.~I.}\ \bibnamefont {Cirac}}, \ and\ \bibinfo {author}
  {\bibfnamefont {M.}~\bibnamefont {Lewenstein}},\ }\href {\doibase
  10.1088/1367-2630/12/11/113037} {\bibfield  {journal} {\bibinfo  {journal}
  {New J. Phys.}\ }\textbf {\bibinfo {volume} {12}},\ \bibinfo {pages} {113037}
  (\bibinfo {year} {2010})}\BibitemShut {NoStop}%
\bibitem [{\citenamefont {Peter}\ \emph {et~al.}(2012)\citenamefont {Peter},
  \citenamefont {M\"uller}, \citenamefont {Wessel},\ and\ \citenamefont
  {B\"uchler}}]{Peter2012}%
  \BibitemOpen
  \bibfield  {author} {\bibinfo {author} {\bibfnamefont {D.}~\bibnamefont
  {Peter}}, \bibinfo {author} {\bibfnamefont {S.}~\bibnamefont {M\"uller}},
  \bibinfo {author} {\bibfnamefont {S.}~\bibnamefont {Wessel}}, \ and\ \bibinfo
  {author} {\bibfnamefont {H.~P.}\ \bibnamefont {B\"uchler}},\ }\href {\doibase
  10.1103/PhysRevLett.109.025303} {\bibfield  {journal} {\bibinfo  {journal}
  {Phys. Rev. Lett.}\ }\textbf {\bibinfo {volume} {109}},\ \bibinfo {pages}
  {025303} (\bibinfo {year} {2012})}\BibitemShut {NoStop}%
\bibitem [{\citenamefont {Nebendahl}\ and\ \citenamefont
  {D\"ur}(2013)}]{Nebendahl2013}%
  \BibitemOpen
  \bibfield  {author} {\bibinfo {author} {\bibfnamefont {V.}~\bibnamefont
  {Nebendahl}}\ and\ \bibinfo {author} {\bibfnamefont {W.}~\bibnamefont
  {D\"ur}},\ }\href {\doibase 10.1103/PhysRevB.87.075413} {\bibfield  {journal}
  {\bibinfo  {journal} {Phys. Rev. B}\ }\textbf {\bibinfo {volume} {87}},\
  \bibinfo {pages} {075413} (\bibinfo {year} {2013})}\BibitemShut {NoStop}%
\bibitem [{\citenamefont {Wall}\ and\ \citenamefont {Carr}(2012)}]{Wall2012}%
  \BibitemOpen
  \bibfield  {author} {\bibinfo {author} {\bibfnamefont {M.~L.}\ \bibnamefont
  {Wall}}\ and\ \bibinfo {author} {\bibfnamefont {L.~D.}\ \bibnamefont
  {Carr}},\ }\href {\doibase 10.1088/1367-2630/14/12/125015} {\bibfield
  {journal} {\bibinfo  {journal} {New J. Phys.}\ }\textbf {\bibinfo {volume}
  {14}},\ \bibinfo {pages} {125015} (\bibinfo {year} {2012})}\BibitemShut
  {NoStop}%
\bibitem [{Note3()}]{Note3}%
  \BibitemOpen
  \bibinfo {note} {When \(L\to \infty \), $f_{k , \alpha } = \protect \frac
  {1}{\protect \text {i}} \left [\protect \mathrm {Li}_{\alpha }\protect
  \tmspace -\thinmuskip {.1667em}(\protect \text {e}^{\protect \text
  {i}k})-\protect \mathrm {Li}_{\alpha }\protect \tmspace -\thinmuskip
  {.1667em}(\protect \text {e}^{-\protect \text {i}k})\right ]$, with $\protect
  \mathrm {Li}_{\alpha }\protect \tmspace -\thinmuskip {.1667em}(z)$ the
  polylogarithmic functions \cite {ancont, ancont3,
  *Abramowitz1964}}\BibitemShut {NoStop}%
\bibitem [{\citenamefont {Hastings}(2008)}]{hastings08}%
  \BibitemOpen
  \bibfield  {author} {\bibinfo {author} {\bibfnamefont {M.~B.}\ \bibnamefont
  {Hastings}},\ }\href {\doibase 10.1088/1742-5468/2008/01/L01001} {\bibfield
  {journal} {\bibinfo  {journal} {J. Stat. Mech.}\ }\textbf {\bibinfo {volume}
  {2008}},\ \bibinfo {pages} {L01001} (\bibinfo {year} {2008})}\BibitemShut
  {NoStop}%
\bibitem [{\citenamefont {Peschel}\ and\ \citenamefont
  {Truong}(1987)}]{Peschel1987}%
  \BibitemOpen
  \bibfield  {author} {\bibinfo {author} {\bibfnamefont {I.}~\bibnamefont
  {Peschel}}\ and\ \bibinfo {author} {\bibfnamefont {T.~T.}\ \bibnamefont
  {Truong}},\ }\href {\doibase 10.1007/BF01307296} {\bibfield  {journal}
  {\bibinfo  {journal} {Z. Phys. B}\ }\textbf {\bibinfo {volume} {69}},\
  \bibinfo {pages} {385} (\bibinfo {year} {1987})}\BibitemShut {NoStop}%
\bibitem [{\citenamefont {Truong}\ and\ \citenamefont
  {I.~Peschel}(1989)}]{Peschel1989}%
  \BibitemOpen
  \bibfield  {author} {\bibinfo {author} {\bibfnamefont {T.~T.}\ \bibnamefont
  {Truong}}\ and\ \bibinfo {author} {\bibfnamefont {I.}~\bibnamefont
  {I.~Peschel}},\ }\href {\doibase 10.1007/BF01313574} {\bibfield  {journal}
  {\bibinfo  {journal} {Z. Phys. B}\ }\textbf {\bibinfo {volume} {75}},\
  \bibinfo {pages} {119} (\bibinfo {year} {1989})}\BibitemShut {NoStop}%
\bibitem [{\citenamefont {Peschel}\ \emph {et~al.}(1999)\citenamefont
  {Peschel}, \citenamefont {Kaulke},\ and\ \citenamefont
  {Legeza}}]{Peschel1999}%
  \BibitemOpen
  \bibfield  {author} {\bibinfo {author} {\bibfnamefont {I.}~\bibnamefont
  {Peschel}}, \bibinfo {author} {\bibfnamefont {M.}~\bibnamefont {Kaulke}}, \
  and\ \bibinfo {author} {\bibfnamefont {O.}~\bibnamefont {Legeza}},\ }\href
  {\doibase 10.1002/(SICI)1521-3889(199902)8:2<153::AID-ANDP153>3.0.CO;2-N}
  {\bibfield  {journal} {\bibinfo  {journal} {Ann. Phys. (Leipzig)}\ }\textbf
  {\bibinfo {volume} {8}},\ \bibinfo {pages} {153} (\bibinfo {year}
  {1999})}\BibitemShut {NoStop}%
\bibitem [{\citenamefont {Peschel}(2012)}]{Peschel2012}%
  \BibitemOpen
  \bibfield  {author} {\bibinfo {author} {\bibfnamefont {I.}~\bibnamefont
  {Peschel}},\ }\href {\doibase 10.1007/s13538-012-0074-1} {\bibfield
  {journal} {\bibinfo  {journal} {Brazilian Journal of Physics}\ }\textbf
  {\bibinfo {volume} {42}},\ \bibinfo {pages} {267} (\bibinfo {year}
  {2012})}\BibitemShut {NoStop}%
\bibitem [{\citenamefont {Eisert}\ \emph {et~al.}(2010)\citenamefont {Eisert},
  \citenamefont {Cramer},\ and\ \citenamefont {Plenio}}]{engl}%
  \BibitemOpen
  \bibfield  {author} {\bibinfo {author} {\bibfnamefont {J.}~\bibnamefont
  {Eisert}}, \bibinfo {author} {\bibfnamefont {M.}~\bibnamefont {Cramer}}, \
  and\ \bibinfo {author} {\bibfnamefont {M.}~\bibnamefont {Plenio}},\ }\href
  {\doibase 10.1103/RevModPhys.82.277} {\bibfield  {journal} {\bibinfo
  {journal} {Rev. Mod. Phys,}\ }\textbf {\bibinfo {volume} {82}},\ \bibinfo
  {pages} {277} (\bibinfo {year} {2010})}\BibitemShut {NoStop}%
\bibitem [{\citenamefont {Amico}\ \emph {et~al.}(2008)\citenamefont {Amico},
  \citenamefont {Fazio}, \citenamefont {Osterloh},\ and\ \citenamefont
  {Vedral}}]{fazio2008}%
  \BibitemOpen
  \bibfield  {author} {\bibinfo {author} {\bibfnamefont {L.}~\bibnamefont
  {Amico}}, \bibinfo {author} {\bibfnamefont {R.}~\bibnamefont {Fazio}},
  \bibinfo {author} {\bibfnamefont {A.}~\bibnamefont {Osterloh}}, \ and\
  \bibinfo {author} {\bibfnamefont {V.}~\bibnamefont {Vedral}},\ }\href
  {\doibase 10.1103/RevModPhys.80.517} {\bibfield  {journal} {\bibinfo
  {journal} {Rev. Mod. Phys.}\ }\textbf {\bibinfo {volume} {80}},\ \bibinfo
  {pages} {517} (\bibinfo {year} {2008})}\BibitemShut {NoStop}%
\bibitem [{\citenamefont {Hastings}\ and\ \citenamefont {Koma}(2006)}]{hast}%
  \BibitemOpen
  \bibfield  {author} {\bibinfo {author} {\bibfnamefont {M.}~\bibnamefont
  {Hastings}}\ and\ \bibinfo {author} {\bibfnamefont {T.}~\bibnamefont
  {Koma}},\ }\href {\doibase 10.1007/s00220-006-0030-4} {\bibfield  {journal}
  {\bibinfo  {journal} {Comm. Math. Phys.}\ }\textbf {\bibinfo {volume}
  {265}},\ \bibinfo {pages} {781} (\bibinfo {year} {2006})}\BibitemShut
  {NoStop}%
\bibitem [{\citenamefont {Holzhey}\ \emph {et~al.}(1994)\citenamefont
  {Holzhey}, \citenamefont {Larsen},\ and\ \citenamefont {Wilczek}}]{holz94}%
  \BibitemOpen
  \bibfield  {author} {\bibinfo {author} {\bibfnamefont {C.}~\bibnamefont
  {Holzhey}}, \bibinfo {author} {\bibfnamefont {F.}~\bibnamefont {Larsen}}, \
  and\ \bibinfo {author} {\bibfnamefont {F.}~\bibnamefont {Wilczek}},\ }\href
  {\doibase 10.1016/0550-3213(94)90402-2} {\bibfield  {journal} {\bibinfo
  {journal} {Nucl. Phys. B}\ }\textbf {\bibinfo {volume} {424}},\ \bibinfo
  {pages} {443} (\bibinfo {year} {1994})}\BibitemShut {NoStop}%
\bibitem [{\citenamefont {Calabrese}\ and\ \citenamefont
  {Cardy}(2004)}]{cal04}%
  \BibitemOpen
  \bibfield  {author} {\bibinfo {author} {\bibfnamefont {P.}~\bibnamefont
  {Calabrese}}\ and\ \bibinfo {author} {\bibfnamefont {J.}~\bibnamefont
  {Cardy}},\ }\href {http://stacks.iop.org/1742-5468/2004/i=06/a=P06002}
  {\bibfield  {journal} {\bibinfo  {journal} {J. Stat. Mech.}\ }\textbf
  {\bibinfo {volume} {2004}},\ \bibinfo {pages} {P06002} (\bibinfo {year}
  {2004})}\BibitemShut {NoStop}%
\bibitem [{Note4()}]{Note4}%
  \BibitemOpen
  \bibinfo {note} {See the Supplemental Material}\BibitemShut {NoStop}%
\bibitem [{\citenamefont {Eisert}\ and\ \citenamefont
  {Osborne}(2006)}]{Eisert2006}%
  \BibitemOpen
  \bibfield  {author} {\bibinfo {author} {\bibfnamefont {J.}~\bibnamefont
  {Eisert}}\ and\ \bibinfo {author} {\bibfnamefont {T.~J.}\ \bibnamefont
  {Osborne}},\ }\href {\doibase 10.1103/PhysRevLett.97.150404} {\bibfield
  {journal} {\bibinfo  {journal} {Phys. Rev. Lett.}\ }\textbf {\bibinfo
  {volume} {97}},\ \bibinfo {pages} {150404} (\bibinfo {year}
  {2006})}\BibitemShut {NoStop}%
\bibitem [{Note5()}]{Note5}%
  \BibitemOpen
  \bibinfo {note} {The low-lying critical finite-size spectrum, we computed,
  has the same degeneracy pattern as the one of the Ising model, for all
  $\alpha $}\BibitemShut {NoStop}%
\bibitem [{\citenamefont {Gradshteyn}\ and\ \citenamefont
  {Ryzhik}(2007)}]{ancont}%
  \BibitemOpen
  \bibfield  {author} {\bibinfo {author} {\bibfnamefont {I.~S.}\ \bibnamefont
  {Gradshteyn}}\ and\ \bibinfo {author} {\bibfnamefont {I.~M.}\ \bibnamefont
  {Ryzhik}},\ }\href@noop {} {\emph {\bibinfo {title} {Tables of Integrals,
  Series, and Products}}}\ (\bibinfo  {publisher} {Academic},\ \bibinfo
  {address} {New York},\ \bibinfo {year} {2007})\BibitemShut {NoStop}%
\bibitem [{\citenamefont {Belavin}\ \emph {et~al.}(1984)\citenamefont
  {Belavin}, \citenamefont {Polyakov},\ and\ \citenamefont
  {Zamolodchikov}}]{Belavin1984}%
  \BibitemOpen
  \bibfield  {author} {\bibinfo {author} {\bibfnamefont {A.~A.}\ \bibnamefont
  {Belavin}}, \bibinfo {author} {\bibfnamefont {A.~M.}\ \bibnamefont
  {Polyakov}}, \ and\ \bibinfo {author} {\bibfnamefont {A.~B.}\ \bibnamefont
  {Zamolodchikov}},\ }\href {\doibase 10.1016/0550-3213(84)90052-X} {\bibfield
  {journal} {\bibinfo  {journal} {Nucl. Phys. B}\ }\textbf {\bibinfo {volume}
  {241}},\ \bibinfo {pages} {333} (\bibinfo {year} {1984})}\BibitemShut
  {NoStop}%
\bibitem [{\citenamefont {Calabrese}\ and\ \citenamefont
  {Cardy}(2005)}]{Calabrese2005}%
  \BibitemOpen
  \bibfield  {author} {\bibinfo {author} {\bibfnamefont {P.}~\bibnamefont
  {Calabrese}}\ and\ \bibinfo {author} {\bibfnamefont {J.}~\bibnamefont
  {Cardy}},\ }\href {http://stacks.iop.org/1742-5468/2005/i=04/a=P04010}
  {\bibfield  {journal} {\bibinfo  {journal} {J. Stat. Mech.}\ }\textbf
  {\bibinfo {volume} {2005}},\ \bibinfo {pages} {P04010} (\bibinfo {year}
  {2005})}\BibitemShut {NoStop}%
\bibitem [{Note6()}]{Note6}%
  \BibitemOpen
  \bibinfo {note} {For, e.g., the Ising model with short-range interactions,
  contributions at $k=0$ in Eq.~\protect \textup {\hbox {\mathsurround \z@
  \protect \normalfont (\ignorespaces \ref {Eq:Corr}\unskip \@@italiccorr )}}
  vanish}\BibitemShut {NoStop}%
\bibitem [{\citenamefont {Lepori}\ \emph {et~al.}(shed)\citenamefont {Lepori}
  \emph {et~al.}}]{Lepori}%
  \BibitemOpen
  \bibfield  {author} {\bibinfo {author} {\bibfnamefont {L.}~\bibnamefont
  {Lepori}} \emph {et~al.},\ }\href@noop {} {} (\bibinfo {year}
  {unpublished})\BibitemShut {NoStop}%
\bibitem [{Note7()}]{Note7}%
  \BibitemOpen
  \bibinfo {note} {This is, for example, at the origin of the behavior of
  $c_{\protect \mathrm {eff}}$, which is different from zero when $1<\alpha <2$
  in the vicinity of the critical line $\mu =-1$}\BibitemShut {NoStop}%
\bibitem [{\citenamefont {Lieb}\ and\ \citenamefont {Robinson}(1972)}]{lieb}%
  \BibitemOpen
  \bibfield  {author} {\bibinfo {author} {\bibfnamefont {E.}~\bibnamefont
  {Lieb}}\ and\ \bibinfo {author} {\bibfnamefont {D.}~\bibnamefont
  {Robinson}},\ }\href {\doibase 10.1007/BF01645779} {\bibfield  {journal}
  {\bibinfo  {journal} {Commun. Math. Phys.}\ }\textbf {\bibinfo {volume}
  {28}},\ \bibinfo {pages} {251} (\bibinfo {year} {1972})}\BibitemShut
  {NoStop}%
\bibitem [{\citenamefont {Lieb}\ \emph {et~al.}(1961)\citenamefont {Lieb},
  \citenamefont {Schultz},\ and\ \citenamefont {Mattis}}]{Lieb1961}%
  \BibitemOpen
  \bibfield  {author} {\bibinfo {author} {\bibfnamefont {E.}~\bibnamefont
  {Lieb}}, \bibinfo {author} {\bibfnamefont {T.}~\bibnamefont {Schultz}}, \
  and\ \bibinfo {author} {\bibfnamefont {D.}~\bibnamefont {Mattis}},\ }\href
  {\doibase 10.1016/0003-4916(61)90115-4} {\bibfield  {journal} {\bibinfo
  {journal} {Annals of Physics}\ }\textbf {\bibinfo {volume} {16}},\ \bibinfo
  {pages} {407} (\bibinfo {year} {1961})}\BibitemShut {NoStop}%
\bibitem [{\citenamefont {Olver}\ \emph {et~al.}(2010)\citenamefont {Olver},
  \citenamefont {Lozier}, \citenamefont {Boisvert},\ and\ \citenamefont
  {Clark}}]{ancont3}%
  \BibitemOpen
  \bibfield  {author} {\bibinfo {author} {\bibfnamefont {F.~W.~J.}\
  \bibnamefont {Olver}}, \bibinfo {author} {\bibfnamefont {D.~W.}\ \bibnamefont
  {Lozier}}, \bibinfo {author} {\bibfnamefont {R.~F.}\ \bibnamefont
  {Boisvert}}, \ and\ \bibinfo {author} {\bibfnamefont {C.~W.}\ \bibnamefont
  {Clark}},\ }\href@noop {} {\emph {\bibinfo {title} {NIST Handbook of
  Mathematical Functions}}}\ (\bibinfo  {publisher} {Cambridge University
  Press},\ \bibinfo {address} {Cambridge, England},\ \bibinfo {year}
  {2010})\BibitemShut {NoStop}%
\bibitem [{\citenamefont {Abramowitz}\ and\ \citenamefont
  {Stegun}(1964)}]{Abramowitz1964}%
  \BibitemOpen
  \bibfield  {author} {\bibinfo {author} {\bibfnamefont {M.}~\bibnamefont
  {Abramowitz}}\ and\ \bibinfo {author} {\bibfnamefont {I.~A.}\ \bibnamefont
  {Stegun}},\ }\href@noop {} {\emph {\bibinfo {title} {Handbook of Mathematical
  Functions}}}\ (\bibinfo  {publisher} {Dover},\ \bibinfo {address} {New
  York},\ \bibinfo {year} {1964})\BibitemShut {NoStop}%
\end{thebibliography}

\begin{thebibliography}{11}%
\makeatletter
\providecommand \@ifxundefined [1]{%
 \@ifx{#1\undefined}
}%
\providecommand \@ifnum [1]{%
 \ifnum #1\expandafter \@firstoftwo
 \else \expandafter \@secondoftwo
 \fi
}%
\providecommand \@ifx [1]{%
 \ifx #1\expandafter \@firstoftwo
 \else \expandafter \@secondoftwo
 \fi
}%
\providecommand \natexlab [1]{#1}%
\providecommand \enquote  [1]{``#1''}%
\providecommand \bibnamefont  [1]{#1}%
\providecommand \bibfnamefont [1]{#1}%
\providecommand \citenamefont [1]{#1}%
\providecommand \href@noop [0]{\@secondoftwo}%
\providecommand \href [0]{\begingroup \@sanitize@url \@href}%
\providecommand \@href[1]{\@@startlink{#1}\@@href}%
\providecommand \@@href[1]{\endgroup#1\@@endlink}%
\providecommand \@sanitize@url [0]{\catcode `\\12\catcode `\$12\catcode
  `\&12\catcode `\#12\catcode `\^12\catcode `\_12\catcode `\%12\relax}%
\providecommand \@@startlink[1]{}%
\providecommand \@@endlink[0]{}%
\providecommand \url  [0]{\begingroup\@sanitize@url \@url }%
\providecommand \@url [1]{\endgroup\@href {#1}{\urlprefix }}%
\providecommand \urlprefix  [0]{URL }%
\providecommand \Eprint [0]{\href }%
\providecommand \doibase [0]{http://dx.doi.org/}%
\providecommand \selectlanguage [0]{\@gobble}%
\providecommand \bibinfo  [0]{\@secondoftwo}%
\providecommand \bibfield  [0]{\@secondoftwo}%
\providecommand \translation [1]{[#1]}%
\providecommand \BibitemOpen [0]{}%
\providecommand \bibitemStop [0]{}%
\providecommand \bibitemNoStop [0]{.\EOS\space}%
\providecommand \EOS [0]{\spacefactor3000\relax}%
\providecommand \BibitemShut  [1]{\csname bibitem#1\endcsname}%
\let\auto@bib@innerbib\@empty
\bibitem [{\citenamefont {Ablowitz}\ and\ \citenamefont
  {Fokas}(2003)}]{Sup_Fokas}%
  \BibitemOpen
  \bibfield  {author} {\bibinfo {author} {\bibfnamefont {M.~J.}\ \bibnamefont
  {Ablowitz}}\ and\ \bibinfo {author} {\bibfnamefont {A.~T.~S.}\ \bibnamefont
  {Fokas}},\ }\href@noop {} {\emph {\bibinfo {title} {Complex Variables
  Introduction and Applications}}}\ (\bibinfo  {publisher} {Cambridge
  University Press},\ \bibinfo {year} {2003})\BibitemShut {NoStop}%
\bibitem [{\citenamefont {Giamarchi}(2004)}]{Sup_GiamarchiBook}%
  \BibitemOpen
  \bibfield  {author} {\bibinfo {author} {\bibfnamefont {T.}~\bibnamefont
  {Giamarchi}},\ }\href@noop {} {\emph {\bibinfo {title} {Quantum Physics in
  One Dimensions}}}\ (\bibinfo  {publisher} {Clarendon Press},\ \bibinfo {year}
  {2004})\BibitemShut {NoStop}%
\bibitem [{\citenamefont {Clark}(2010)}]{Sup_ancont3}%
  \BibitemOpen
  \bibfield  {author} {\bibinfo {author} {\bibfnamefont {C.~W.}\ \bibnamefont
  {Clark}},\ }\href@noop {} {\emph {\bibinfo {title} {NIST Handbook of
  Mathematical Functions}}}\ (\bibinfo  {publisher} {Cambridge University
  Press},\ \bibinfo {year} {2010})\BibitemShut {NoStop}%
\bibitem [{\citenamefont {Gradshteyn}\ and\ \citenamefont
  {Ryzhik}(2007)}]{Sup_ancont}%
  \BibitemOpen
  \bibfield  {author} {\bibinfo {author} {\bibfnamefont {I.~S.}\ \bibnamefont
  {Gradshteyn}}\ and\ \bibinfo {author} {\bibfnamefont {I.~M.}\ \bibnamefont
  {Ryzhik}},\ }\href@noop {} {\emph {\bibinfo {title} {Tables of Integrals,
  Series, and Products}}}\ (\bibinfo  {publisher} {Academic Press},\ \bibinfo
  {year} {2007})\BibitemShut {NoStop}%
\bibitem [{\citenamefont {Lieb}\ \emph {et~al.}(1961)\citenamefont {Lieb},
  \citenamefont {Schultz},\ and\ \citenamefont {Mattis}}]{Sup_Lieb1961}%
  \BibitemOpen
  \bibfield  {author} {\bibinfo {author} {\bibfnamefont {E.}~\bibnamefont
  {Lieb}}, \bibinfo {author} {\bibfnamefont {T.}~\bibnamefont {Schultz}}, \
  and\ \bibinfo {author} {\bibfnamefont {D.}~\bibnamefont {Mattis}},\ }\href
  {\doibase 10.1016/0003-4916(61)90115-4} {\bibfield  {journal} {\bibinfo
  {journal} {Annals of Physics}\ }\textbf {\bibinfo {volume} {16}},\ \bibinfo
  {pages} {407} (\bibinfo {year} {1961})}\BibitemShut {NoStop}%
\bibitem [{\citenamefont {Blaizot}\ and\ \citenamefont
  {Ripka}(1986)}]{Sup_Ripka}%
  \BibitemOpen
  \bibfield  {author} {\bibinfo {author} {\bibfnamefont {J.~P.}\ \bibnamefont
  {Blaizot}}\ and\ \bibinfo {author} {\bibfnamefont {G.}~\bibnamefont
  {Ripka}},\ }\href@noop {} {\emph {\bibinfo {title} {Quantum Theory of Finite
  Systems}}}\ (\bibinfo  {publisher} {MIT Press},\ \bibinfo {year}
  {1986})\BibitemShut {NoStop}%
\bibitem [{\citenamefont {Peschel}\ and\ \citenamefont
  {Truong}(1987)}]{Sup_Peschel1987}%
  \BibitemOpen
  \bibfield  {author} {\bibinfo {author} {\bibfnamefont {I.}~\bibnamefont
  {Peschel}}\ and\ \bibinfo {author} {\bibfnamefont {T.~T.}\ \bibnamefont
  {Truong}},\ }\href {\doibase 10.1007/BF01307296} {\bibfield  {journal}
  {\bibinfo  {journal} {Z. Phys. B}\ }\textbf {\bibinfo {volume} {69}},\
  \bibinfo {pages} {385} (\bibinfo {year} {1987})}\BibitemShut {NoStop}%
\bibitem [{\citenamefont {Truong}\ and\ \citenamefont
  {I.~Peschel}(1989)}]{Sup_Peschel1989}%
  \BibitemOpen
  \bibfield  {author} {\bibinfo {author} {\bibfnamefont {T.~T.}\ \bibnamefont
  {Truong}}\ and\ \bibinfo {author} {\bibfnamefont {I.}~\bibnamefont
  {I.~Peschel}},\ }\href {\doibase 10.1007/BF01313574} {\bibfield  {journal}
  {\bibinfo  {journal} {Z. Phys. B}\ }\textbf {\bibinfo {volume} {75}},\
  \bibinfo {pages} {119} (\bibinfo {year} {1989})}\BibitemShut {NoStop}%
\bibitem [{\citenamefont {Peschel}\ \emph {et~al.}(1999)\citenamefont
  {Peschel}, \citenamefont {Kaulke},\ and\ \citenamefont
  {Legeza}}]{Sup_Peschel1999}%
  \BibitemOpen
  \bibfield  {author} {\bibinfo {author} {\bibfnamefont {I.}~\bibnamefont
  {Peschel}}, \bibinfo {author} {\bibfnamefont {M.}~\bibnamefont {Kaulke}}, \
  and\ \bibinfo {author} {\bibfnamefont {O.}~\bibnamefont {Legeza}},\ }\href
  {\doibase 10.1002/(SICI)1521-3889(199902)8:2<153::AID-ANDP153>3.0.CO;2-N}
  {\bibfield  {journal} {\bibinfo  {journal} {Ann. Phys. (Leipzig)}\ }\textbf
  {\bibinfo {volume} {8}},\ \bibinfo {pages} {153} (\bibinfo {year}
  {1999})}\BibitemShut {NoStop}%
\bibitem [{\citenamefont {Peschel}(2003)}]{Sup_Peschel2002}%
  \BibitemOpen
  \bibfield  {author} {\bibinfo {author} {\bibfnamefont {I.}~\bibnamefont
  {Peschel}},\ }\href {\doibase 10.1088/0305-4470/36/14/101} {\bibfield
  {journal} {\bibinfo  {journal} {J. Phys. A}\ }\textbf {\bibinfo {volume}
  {36}},\ \bibinfo {pages} {L205} (\bibinfo {year} {2003})}\BibitemShut
  {NoStop}%
\bibitem [{\citenamefont {Peschel}(2012)}]{Sup_Peschel2012}%
  \BibitemOpen
  \bibfield  {author} {\bibinfo {author} {\bibfnamefont {I.}~\bibnamefont
  {Peschel}},\ }\href {\doibase 10.1007/s13538-012-0074-1} {\bibfield
  {journal} {\bibinfo  {journal} {Brazilian Journal of Physics}\ }\textbf
  {\bibinfo {volume} {42}},\ \bibinfo {pages} {267} (\bibinfo {year}
  {2012})}\BibitemShut {NoStop}%
\end{thebibliography}
%

\end{document}